\documentclass[aps,prd,groupedaddress,nofootinbib]{revtex4}
\def\ga{\mathrel{\raise.3ex\hbox{$>$\kern-.75em\lower1ex\hbox{$\sim$}}}}
\def\la{\mathrel{\raise.3ex\hbox{$<$\kern-.75em\lower1ex\hbox{$\sim$}}}}

\def\be {\beta}
\def\al {\alpha}

\newcommand{\lam}{\lambda}
\def\lsim{\mathrel{\rlap{\lower4pt\hbox{\hskip1pt$\sim$}}
    \raise1pt\hbox{$<$}}}                % less than or approx. symbol
\def\gsim{\mathrel{\rlap{\lower4pt\hbox{\hskip1pt$\sim$}}
    \raise1pt\hbox{$>$}}}                % greater than or approx. symbol

\usepackage{amsmath}
\usepackage{amssymb}
\usepackage{bbm}
\usepackage{epsfig}
\begin{document}
\begin{flushright}
SHEP-09-09
\end{flushright}
\title{Double Neutral Higgs production in the Two-Higgs
doublet model at the LHC}
\author{Abdesslam Arhrib$^{1,2,3}$, Rachid Benbrik$^{3,4,5}$,
Chuan-Hung  Chen$^{4,5}$, Renato Guedes$^{6}$ and Rui
Santos$^{6}$}
\affiliation{$^1$ National Center for Theoretical Physics,
National Taiwan University, Taipei, Taiwan 10617, R.O.C}
\affiliation{$^2$ D\'epartement de Math\'ematique, Facult\'e des
Sciences and Techniques, Universit\'e Abdelmalek Essa\^adi, B.
416, Tangier, Morocco. }
\affiliation{$^3$LPHEA, Facult\'e des
Sciences-Semlalia, B.P. 2390 Marrakesh, Morocco.}
\affiliation{$^4$ Department of Physics, National Cheng Kung
  University, Taiwan 701, Taiwan.}
\affiliation{$^5$ National Center for Theoretical Physics,
 Taiwan 701}
\affiliation{$^6$ NExT Institute and School of Physics and
Astronomy, University of Southampton Highfield, Southampton SO17
1BJ, UK.}
\date{\today}

\begin{abstract}
\noindent In this work we study the production and decay of two CP-even Higgs bosons
in the general CP-conserving two-Higgs doublet model at the LHC. We also study
the limiting case of the decoupling scenario. For each Yukawa version of the
model, we look for the region of the parameter space where a signal would be
seen at the LHC. Taking into account theoretical and experimental constraints
on the two Higgs doublet model parameters,
we show that the cross section can be two orders of magnitude above the
corresponding Standard Model cross section. We have also studied in detail the decay profile of the Higgs bosons and showed that interesting signatures may emerge for some particular values of the parameters. In some scenarios,
specific triple Higgs could be measured.
\end{abstract}


\maketitle
\section{Introduction}
CERN's Large Hadron Collider (LHC) is due to start operation in 2009 with a
center of mass energy of 10 $TeV$. Therefore, testing the spontaneous symmetry
breaking (SSB) mechanism and the search for the scalar particle or particles
associated to that process is most probably postponed for at least another
year. On the other hand, the Tevatron~\cite{Bernardi:2008ee} has recently
excluded a 160-170 $GeV$ Standard Model (SM) Higgs which decays to a pair of W
gauge bosons. The mass region probed will obviously grow as data is being
collected and a Higgs boson can still be discovered at the Tevatron.

If a Higgs boson is found, the next task will be to understand the underlying
mechanism that gives mass to all known particles. Characterizing the Higgs
potential is the first step towards inferring the nature of the SSB
mechanism. There are several alternatives to the SM that can be tested at the
LHC. The most popular are the Minimal Supersymmetric Model (MSSM) or some
of its variants like two Higgs Doublet Model (2HDM),
little Higgs models and extra dimensions models among others.
It is well known that the CP-conserving 2HDM has in its spectrum 2 CP-even
scalars, $h^0$ and $H^0$, one CP-odd scalar $A^0$ and a pair of charged Higgs bosons, $H^\pm$. Regarding the Higgs sector, the 2HDM is a clear candidate to be probed at the LHC because it is the minimal version of the SM that gives rise to charged Higgs bosons while being consistent with all available experimental data. The difference between all these models can in principle be investigated by examining the processes involving scalar triple and quartic vertices.

In the SM, the potential can be studied in processes which have at least two Higgs in final state. The process $gg \to h^0 h^0$ was studied in the SM in~\cite{Glover:1987nx,Dicus:1987ic}, in the SM with a sequential fourth generation in \cite{Kribs:2007nz} and in the MSSM in~\cite{Plehn:1996wb}. The SM cross section is of the order of 20 $fb$ for a Higgs mass of 115 $GeV$ and it drops steeply with increasing Higgs mass as a consequence of the fast fall–off of the gluon-gluon luminosity. A sequential fourth generation can enhance the double Higgs production rate to $\approx 300$ fb for a Higgs mass of 115 $GeV$. In the MSSM, the triple coupling can be probed in a wide range of Higgs masses whenever the resonant decay $H^0 \to h^0 h^0$ is possible. The measurement will be easier for small $\tan \beta$, where the trilinear coupling is larger and the
heavy Higgs mass is less than twice the light Higgs mass in major parts of the parameter space. More detailed studies at parton level were then performed for
the SM in~\cite{Baur:2002qd,Baur:2003gp,Baur:2003gpa} and for the MSSM in~\cite{Dai:1995cb,Dai:1996rn}. Other processes such as vector boson fusion have been proposed for the 2HDM and shown to be rather powerful in reconstructing the triple Higgs self coupling \cite{Moretti:2007ca}.

The process $pp \to h^0 h^0$ is also possible in the general 2HDM. In order to be able to distinguish between models, such as 2HDM, MSSM, sequential fourth generation and little Higgs models, it is mandatory to perform a detailed study for the double Higgs production within the general 2HDM. In a previous study, we have considered double Higgs production at hadron colliders in the 2HDM fermiophobic limit~\cite{Arhrib:2008pw} and we have shown that information on the triple Higgs couplings could be extracted. In this work, we extend the above study to the general 2HDM. We will consider all CP-even double Higgs production, that is, $pp \to h^0 h^0 , h^0 H^0 , H^0 H^0$. We will present the regions of the parameter space of a general CP-conserving 2HDM where this process is large enough to be probed at the LHC taking into account the various  experimental constraints on the 2HDM parameters as well as the tree-level unitarity and vacuum stability constraints.  We will show that in some cases, even after imposing all constraints, cross sections are still large enough to allow for a determination of 2HDM triple Higgs couplings. In other cases only a deviation from the SM can be measured and other processes would be needed to complete the picture. We will give particular attention to the decoupling limit where the lightest Higgs mimics the SM Higgs and look for the possible non-decoupling effects.

This paper is organized as follows. In the next section we review the 2HDM potential. We then present all experimental and theoretical constraints that bound the 2HDM parameter space. In Section~\ref{sec:numerical} we present our numerical results. In Section~\ref{sec:signatures} we discuss the final states in the different 2HDM scenarios. Finally, our findings are summarized and discussed in Section~\ref{sec:summary}.

\section{The CP-conserving 2HDM} The most general two-Higgs doublet model has 14 independent parameters. It violates CP explicitly and it can break electric charge. Even if explicit CP-violating interactions are not present in the potential, spontaneous CP breaking can still occur. One way to force the CP minimum not to exist is to impose the exact $Z_2$ discrete symmetry $\Phi_1 \to \Phi_1$, $\Phi_2 \to - \Phi_2$~\cite{Velhinho:1994vh}. This symmetry can be softly broken by dimension two terms we define as $[m_{12}^2\Phi_1^\dagger\Phi_2+{\rm h.c.}]$. In this case, we can build either a CP-conserving theory or one that breaks CP spontaneously. In this work the vacuum structure is chosen so that the potential is CP-conserving. The theory build on a minimum that does not break CP-invariance nor electric charge was shown to be stable at tree level~\cite{vacstab1,vacstab2}. Under these constraints, the most general renormalizable potential which is invariant under $SU(2) \otimes
 U(1)$ can be written as
\begin{eqnarray}
V(\Phi_1,\Phi_2) &=& m^2_1 \Phi^{\dagger}_1\Phi_1+m^2_2
\Phi^{\dagger}_2\Phi_2 + (m^2_{12} \Phi^{\dagger}_1\Phi_2+{\rm
h.c}) +\frac{1}{2} \lam_1 (\Phi^{\dagger}_1\Phi_1)^2 +\frac{1}{2}
\lam_2 (\Phi^{\dagger}_2\Phi_2)^2\nonumber \\ &+& \lam_3
(\Phi^{\dagger}_1\Phi_1)(\Phi^{\dagger}_2\Phi_2) + \lam_4
(\Phi^{\dagger}_1\Phi_2)(\Phi^{\dagger}_2\Phi_1) + \frac{1}{2}
\lam_5[(\Phi^{\dagger}_1\Phi_2)^2+{\rm h.c.}] ~, \label{higgspot}
\end{eqnarray}
where $\Phi_i$, $i=1,2$ are complex $SU(2)$ doublets with 4
degrees of freedom each and all $m_{i}^2$, $\lambda_i$ and $m_{12}^2$ are real
which follows from the hermiticity of the potential. From the initial eight
degrees of freedom, if the $SU(2)$ symmetry is broken, we end up with two
CP-even Higgs states usually denoted by $h^0$ and $H^0$, one CP-odd state,
$A^0$, two charged Higgs boson, $H^{\pm}$ and three Goldstone bosons.

The potential in Eq.~(\ref{higgspot}) has a total of 10 parameters if one includes the vacuum expectation values. In a CP-conserving minimum there are two minimization conditions that can be used to fix the tree-level value of the parameters $m_1^2$ and $m_2^2$. The combination $v^2=v_1^2 + v_2^2$ is fixed as usual by the electroweak breaking scale through $v^2=(2\sqrt{2} G_F)^{-1}$.  We are thus left with 7 independent parameters, namely $(\lambda_i)_{i=1,\ldots,5}$, $m_{12}^2$, and $\tan\beta \equiv v_2/v_1$.  Equivalently, we can take instead the set, $m_{h^0}$, $m_{H^0}$, $m_{A^0}$, $m_{H^\pm}$, $\tan\beta$, $\alpha$ and $m_{12}^2$, as the 7 independent parameters. The angle $\beta$ is the rotation angle from the group eigenstates to the mass eigenstates in the CP-odd and charged sector. The angle $\alpha$ is the corresponding rotation angle for the CP-even sector. The parameter $m_{12}$ is a measure of how the discrete symmetry is broken. As stated previously, the potential with $m_{12}=0$ has an exact $Z_2$ symmetry and is always CP-conserving.

From the above potential, Eq.(\ref{higgspot}), we can derive the triple Higgs couplings needed for the present study
as a function of the 2HDM parameters $m_{h^0}$, $m_{H^0}$, $m_{A^0}$, $m_{H^\pm}$, $\tan\beta$, $\alpha$ and $m_{12}^2$. They are the same for all different Yukawa versions of the 2HDM and are given by:
\begin{eqnarray}
\lambda_{h^0h^0h^0}^{2HDM} &=& \frac{-3e}{m_W s_W s^2_{2\be}}\bigg[(c_\be c^3_\al - s_\be
  s^3_\al)s_{2\be} m^2_{h^0} + c^2_{\be-\al} c_{\be + \al} m^2_{12}\bigg]
\label{lll}\\
\lambda_{H^0H^0H^0}^{2HDM} &=& \frac{-3e}{m_W s_W s^2_{2\be}}\bigg[(c_\be c^3_\al - s_\be
  s^3_\al)s_{2\be} m^2_{H^0} + s^2_{\be-\al} s_{\be + \al} m^2_{12}\bigg]
\label{hhh}\\
\lambda_{H^0h^0h^0}^{2HDM} &=& -\frac{1}{2}\frac{e c_{\be-\al}}{m_W s_W s^2_{2\be}}\bigg[
  (2 m^2_{h^0} + m^2_{H^0}) s_{2\al} s_{2\be} + (3 s_{2\al}-s_{2\be})
  m^2_{12}\bigg]\label{hll} \\
\lambda_{H^0H^0h^0}^{2HDM} &=& \frac{1}{2}\frac{e s_{\be-\al}}{m_W s_W s^2_{2\be}}\bigg[
  (m^2_{h^0} + 2 m^2_{H^0}) s_{2\al} s_{2\be} + (3 s_{2\al}+s_{2\be})
  m^2_{12}\bigg] \label{hhl}
\end{eqnarray}
\begin{eqnarray}
\lambda_{h^0h^0h^0}^{SM} & = & \frac{-3em_{h^0}^2}{2 m_W s_W}\label{hhhsm}
\end{eqnarray}
where $\lambda_{h^0h^0h^0}^{SM}$ is the triple Higgs SM coupling, $e$ is the electric charge, $m_W$ is the $W$ boson mass, $s_W$ is the $\sin$ of the Weinberg angle and $s$ and $c$ were used as shorthand notation for $\sin$ and $\cos$.
As we can see from Eqs.~(\ref{lll})-(\ref{hhl}),
all triple Higgs couplings have a quadratic dependence
of the physical masses of the fields present in the vertex and on the soft breaking term $m_{12}^2$. These couplings also depend strongly on $\tan\beta$ and $\alpha$.

The Yukawa Lagrangian is a straightforward generalization of the SM one. Its most general form in the 2HDM is
\begin{equation}
{\cal {L}}_{Y}\, = \, G_{ij}^k \left( \bar{u} \quad \bar{d} \right)_{L}^i \,
\phi_k \, d_{R}^j + \, \tilde{G}_{ij}^k \left( \bar{u} \quad \bar{d} \right)_{L}^i \,
\tilde{\phi}_k \, u_{R}^j + h.c.
\end{equation}
where the \(G_{ij}^k\) and \(\tilde{G}_{ij}^k\) are arbitrary constants, $i$ and $j$ are generation indices and $k$ is the doublet number. However, tree-level flavor changing neutral currents (FCNCs) are severely constrained by experiment. The symmetry imposed to the potential can be naturally extended to the Yukawa Lagrangian to guarantee that FCNCs are not present. It suffices that fermions of a given electric charge couple to no more than one Higgs doublet~\cite{Glashow}. This can be accomplished naturally by imposing on all fields appropriate
discrete symmetries that forbid the unwanted FCNC couplings. There are essentially four ways of doing this~\cite{barger} and so there are four variations of the model. We define as Type I the model where only the doublet $\phi_2$ couples to all fermions; Type II is the model where $\phi_2$ couples to up-type quarks and $\phi_1$ couples to down-type quarks and leptons; in a Type III model $\phi_2$ couples to all quarks and $\phi_1$ couples to all leptons; a Type IV model is instead built such that $\phi_2$ couples to up-type quarks and to leptons and $\phi_1$ couples to down-type quarks. Models III and IV have been explored in a number of papers~\cite{L2HDM1} and more recently  with a renewed interest in~\cite{L2HDM2}. We present all Yukawa coupling in appendix A. Regarding the Higgs couplings to the quarks, the models can be grouped in pairs like (I, IV) and (II, III) as the only difference between these two groups is in the couplings to the leptons. Moreover, the way we have chosen to build the Yukawa Lagrangian is such that up type quarks have the same couplings to the Higgs bosons in all four models. Down type quarks have different couplings in the two groups defined above, (I, IV) and (II, III), as can be seen in the appendix. In the large $\tan\beta$ limit, in 2HDM-(II, III), the Higgs coupling to a pair of down quarks is enhanced by a factor of $1/\cos\beta \approx \tan\beta$ while for 2HDM-(I, IV) there is no such enhancement. The processes described in this work of neutral Higgs pair production are either one-loop and proceed via quark loops or tree-level in the case of $b \bar{b}$ fusion. We will show that the theoretical constraints force $\tan \beta$ to be small, let us say smaller than 5 (we will come back to this point later). Because the Yukawa couplings are proportional to the quark masses and some function of the angles, for such small values of $\tan \beta$ the bottom loop can not compete with the top loop and the results for $gg \to h^0 h^0$ proved to be almost Yukawa-model independent. We will show that the only difference comes from higher order corrections that enter the calculation via the width of the Higgs bosons.

\section{Theoretical and experimental bounds}
\label{sec:bounds}
The parameter space of the scalar potential of the 2HDM is reduced both by theoretical constraints as well as by the results of experimental searches. From the theoretical constraints which the 2HDM is subjected to, the most important are the ones that insure tree-level unitarity and vacuum stability of the theory. We also force the potential to be perturbative by requiring that all quartic couplings of the scalar potential, Eq.~(\ref{higgspot}), obey $|\lambda_i| \leq 8 \pi$ for all $i$. For the vacuum stability conditions that ensure that the potential is bounded from below, we use those from \cite{vac1}, which are given by
\begin{eqnarray}
\nonumber
& \lambda_1  > 0\;,\quad\quad \lambda_2 > 0\;,
\nonumber\\
& \sqrt{\lambda_1\lambda_2 }
+ \lambda_{3}  + {\rm{min}}
\left( 0 , \lambda_{4}-|\lambda_{5}|
 \right) >0  \, \, . \label{vac}
\end{eqnarray}
The most restrictive theoretical bound comes from the full set of unitarity constraints \cite{unit1,abdesunit} established using the high energy approximation as well as the equivalence theorem and which can be written as
\begin{eqnarray}
  |a_{\pm}|,  |b_{\pm}|,  |c_{\pm}|,  |d_{\pm}|,
   |e_{1,2}^{}|,  |f_{\pm}|,  |g_{1,2}^{}|  < 8 \pi \label{unita}
\end{eqnarray}
with
\begin{eqnarray}
a_{\pm}^{} &=&
 \frac{3}{2} \left\{
  (\lambda_1 + \lambda_2) \pm
   \sqrt{ (\lambda_1-\lambda_2)^2 + \frac{4}{9} (2\lambda_3+\lambda_4)^2}
  \right\},  \\
b_{\pm}^{} &=&
 \frac{1}{2} \left\{
   (\lambda_1 + \lambda_2) \pm
   \sqrt{ (\lambda_1-\lambda_2)^2 +4 \lambda_4^2}
  \right\},  \\
c_{\pm}^{} &=& d_{\pm}^{}=
 \frac{1}{2} \left\{
   (\lambda_1 + \lambda_2) \pm
   \sqrt{(\lambda_1-\lambda_2)^2 +4 \lambda_5^2}
  \right\},  \\
e_1 &=&    \left(
    \lambda_3 + 2 \lambda_4 - 3 \lambda_5 \right) \qquad , \qquad
e_2 =    \left(
    \lambda_3 - \lambda_5 \right), \\
f_+ &=&    \left(
    \lambda_3 + 2 \lambda_4 + 3 \lambda_5 \right) \qquad , \qquad
f_- =    \left(
    \lambda_3 + \lambda_5 \right), \\
g_1 &=& g_2 =    \left(
    \lambda_3 + \lambda_4 \right).
\end{eqnarray}

The 2HDM parameters are also constrained by direct experimental searches and by precision experimental data. First, the extra contributions to the $\delta \rho$ parameter from the extra Higgs scalars \cite{Rhoparam} should not exceed the current limits from precision measurements \cite{pdg4}: $ |\delta\rho| \la 10^{-3}$. Such an extra contribution to $\delta\rho$ vanishes in the limit $m_{H^\pm}=m_{A^0}$. To ensure that  $\delta\rho$ will be within the allowed range we allow only a small splitting between $m_{H^\pm}$ and $m_{A^0}$. Second, it has been shown in Ref.~\cite{Oslandk} that data from $B\to X_s \gamma$ impose a lower limit of $m_{H^\pm} \ga 290$\,GeV in Models type II and III. This constraint no longer applies in the case of 2HDM-I and 2HDM-IV. Third, values of $\tan \beta$ smaller than $\approx 1$ are disallowed both by the constraints coming from $Z \rightarrow b \bar{b}$ and from $B_q \bar{B_q}$ mixing~\cite{Oslandk} for all Yukawa versions of the model. On the other hand, because $\tan \beta$ can not be too large due to the theoretical constraints, except in some very specific scenarios, there is no need to discuss experimental constraints for very large $\tan \beta$. Finally, searches for Higgs bosons at colliders directly constrains their masses. The constraints on neutral Higgs bosons depend on their decay modes. The scenario where the light CP-even Higgs $h^0$ is fermiophobic, has been discussed in \cite{Arhrib:2008pw}. If the neutral Higgs bosons decay mainly into fermions, the OPAL, DELPHI and L3 collaborations have set a limit on the $h^0$ and $A^0$ masses of the 2HDM \cite{opal,L3,delphi}. OPAL concluded that the regions $1\la m_{h^0} \la 55$ GeV and $3\la m_{A^0} \la 63$ GeV are excluded at 95\% CL independently of the values of $\alpha$ and $\tan\beta$~\cite{opal}. It should be noted however that this is true as long as we are away from the limit $\alpha - \beta \approx 0$. In this limit, the cross section for the process  $e^+ e^- \to Z \to Zh^0$ vanishes in the 2HDM and the above bound no longer applies. L3 set a lower limit on the CP-even mass of the order 110.3 GeV \cite{L3} if the Higgs-strahlung cross section is the SM one and the Higgs decays hadronically. The DELPHI collaboration also studied the decay $h^0 \to A^0 A^0$ in $e^+e^-\to h^0 Z$ and $e^+e^-\to h^0 A^0$ production and large portions of the $(m_{h^0},m_{A^0})$ plane were excluded depending on the 2HDM suppression factor that enters the  $e^+e^-\to h^0 Z$ and $e^+e^-\to h^0 A^0 $ cross sections \cite{delphi} with respect to the SM. In what follows, we will assume that all Higgs masses are above 100 GeV unless the ratio of couplings $g_{ZZh^0}^{2HDM}/g_{ZZh}^{SM}$ is suppressed down to $\sin (\beta - \alpha) \approx 0.1$ when no bound to $m_{h^0}$ applies.

\section{Results}
\label{sec:numerical}
\subsection{The calculation}
\begin{figure}[h!]
  \begin{center}
    \epsfig{file=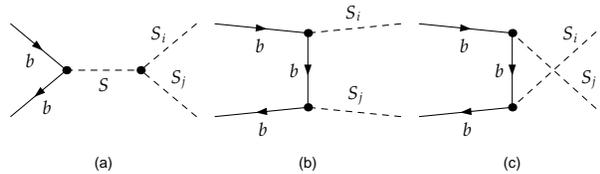,width=12cm}
%\vspace{-1.cm}
    \caption{Tree-level contribution to $b\bar{b} \rightarrow S_i S_j$}
    \label{fig:fig_bornbbhh}
  \end{center}
\end{figure}

The double CP-even Higgs production processes $pp \rightarrow S_iS_j$ ($S_{i,j}=h^0, H^0$) have tree level contributions from $q \bar{q} \rightarrow S_iS_j$ with s-channel Higgs exchange \footnote{Bose Symmetry forbids the couplings $Zh^0h^0$, $Zh^0H^0$ and $ZH^0H^0$.}, t-channel quark exchange and one loop contributions from $gg \rightarrow S_iS_j$. The most relevant contribution for process $q \bar{q} \rightarrow S_iS_j$ is the one where $q=b$ because the Higgs couples proportionally to the quark masses. The tree level diagrams for double Higgs production via b-quark fusion are shown in Fig.~\ref{fig:fig_bornbbhh}.

\begin{figure}[h]
%\vspace{-1.cm}
\begin{center}
\epsfig{file=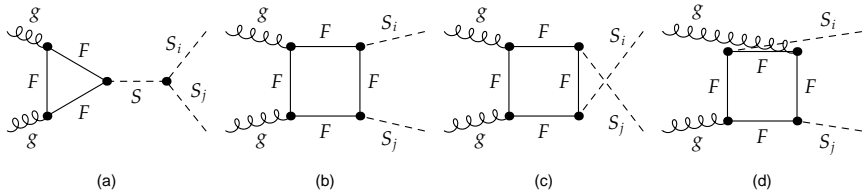,width=12cm}
%\vspace{-5.cm}
\caption{Triangle and box contributions to $gg \rightarrow S_iS_j$,
where $S=h^0,H^0$ and $F$ is a generic quark.}
\label{fig:fig_gghh}
\end{center}
\end{figure}
The process $gg \rightarrow S_i S_j$ occurs only at the
one-loop level and the four types of diagrams that participate in the process are depicted in Fig.~\ref{fig:fig_gghh}. The triangle diagrams, as shown in Fig.~\ref{fig:fig_gghh} (a), represent the generic contribution to the process with a fermion $F$ which can be either an up or a down quark. The second type of diagrams are the box contributions which come in three different topologies, with virtual up and down quarks being exchanged. Comparing the tree-level with the loop diagrams it is easy to understand that no complementarity is expected from these two contributions - if a given set of parameters originate a small value of $\sigma ( b \bar{b} \rightarrow S_i S_j)$ it will forcibly lead to a small $ \sigma ( gg \rightarrow S_i S_j) $ as well. We have explicitly checked that the tree-level contribution is  always well below the loop contribution - usually less than $5 \, \%$. All other quark fusion processes are negligible when compared to $b \bar{b} \rightarrow S_i S_j
 $.

The tree and one-loop amplitudes were generated and calculated with the packages FeynArts \cite{feynarts} and FormCalc \cite{formcalc}. The scalar integrals were evaluated with  LoopTools \cite{looptools}. We have included the total width of the scalar particle $S_i = h^0, H^0$ in the calculation of the corresponding amplitude. Even if the dominant contribution to the one-loop amplitude comes from top and bottom diagrams, in our numerical results we include all quarks in the loops. In models I and IV the largest contribution is always the one with top quarks in the loops. In the remaining two Yukawa models the situation is not as clear - for instance, in the triangle loop diagram with a virtual light Higgs, $S=h^0$, when $m_t \approx \tan \al \, \tan \be \, m_b$ the contribution from the bottom quark triangle loop is larger than the one with the top quark loop. However, the same is not true for the triangle diagram with the virtual $H^0$. We will show that because $\tan \beta
 $ has to be small, the difference between models can only arises mainly from the $\alpha$ angle dependence which is in fact a combined dependence of the two CP-even masses, $m_{12}$, $\tan \beta$ and $\sin \alpha$.

We will start our discussion with the very interesting scenario of the decoupling limit, where the light Higgs resembles the SM one. We will then move to the study of the more general case and try to identify the regions of the 2HDM parameter space that can be probed in a more general case.

\subsection{The decoupling limit}
The decoupling limit of the 2HDM is usually defined as the limiting case where all scalar masses except the lighter CP-even Higgs become infinite and the effective theory is just the SM with one doublet (see \cite{Gunion:2002zf} for a discussion). In this case, the CP-even $h^0$ is the lightest scalar particle while the other Higgs particles, $H^0$, $A^0$ and $H^{\pm}$ are heavy and mass-degenerate. Using purely algebraic arguments at the tree level, one can derive that the main consequences of the decoupling limit is that because $\cos(\beta-\alpha)\to 0$, the CP-even $h^0$ of the 2HDM and the SM Higgs $h_{SM}$ have  similar tree level couplings to the gauge bosons and to the fermions \cite{Gunion:2002zf,okada}. Obviously, the decoupling limit does not rigorously apply to the more realistic world where the particle masses are finite. Actually, one may consider a less rigorous scenario, labeled as the decoupling regime ~\cite{Gunion:2002zf}, where the heavy Higgs particles have masses much larger than the $Z$ boson mass and may escape detection in the planned experiments.

Several studies have been carried out looking for non-decoupling effects in Higgs boson decays and Higgs self-interactions. Large loop effects in $h^0\to \gamma \gamma$, $h^0\to \gamma Z$ and $h^0 \to b\bar{b}$ have been pointed out for the 2HDM~\cite{maria,indirect} and may give indirect information on Higgs masses and the involved triple Higgs couplings  such as  $h^0H^+H^-$, $h^0H^0H^0$, $h^0A^0A^0$ and $h^0h^0h^0$. The non-decoupling contributions to the triple Higgs self coupling $h^0h^0h^0$  has been investigated in the 2HDM in Ref.~\cite{okada}, revealing large non-decoupling effects in the 2HDM. In this section we will show that the large non-decoupling effects in the $h^0h^0h^0$ coupling will modify the double Higgs pair production $pp \to h^0h^0$ cross section and make it larger than its SM counterpart. We will focus on the scenario where all Higgs particles of the 2HDM, except for the lightest CP-even Higgs, are heavy and can escape detection at the early stages of next generation of colliders.

It is easy to check that in the decoupling limit the relation $\beta-\alpha\to \pi/2$ holds; hence, the triple coupling $h^0h^0h^0$ given in Eq.~(\ref{lll}) is reduced to its SM value $g_{h^0h^0h^0}=- 3 m_{h^0}^2/v$. Furthermore, we have $g_{H^0h^0h^0} \to 0$ at tree level and so the $gg \to h^0h^0$ and $b\bar{b} \to h^0h^0$ amplitudes reduce
exactly to the SM results. As shown in~\cite{okada}, the one-loop leading contributions from all heavy Higgs boson loops and also from top quark loops to the effective $h^0h^0h^0$ coupling, can be written as
\begin{eqnarray}
 \lambda_{hhh}^{eff}(2HDM) \!\!&=&\!\! - \frac{3 m_{h^0}^2}{v}
      \left\{ 1
     + \frac{m_{H^0}^4}{12 \pi^2 m_{h^0}^2 v^2}
     \left(1 + \frac{M^2}{m_{H^0}^2}\right)^3
     + \frac{m_{A^0}^4}{12 \pi^2 m_{h^0}^2 v^2}
     \left(1 + \frac{M^2}{m_{A^0}^2}\right)^3 \right.\nonumber\\
&&\left.
\!\!\!\!\!\!\!\!\!\!\!\!\!\!\!\!\!\!\!\!\!\!\!\!\!\!\!\!\!\!
\!\!\!\!\!\!\!\!\!\!\!\!\!\!\!\!\!\!
    + \frac{m_{H^\pm}^4}{6 \pi^2 m_{h^0}^2 v^2}
    \left(1 + \frac{M^2}{m_{H^\pm}^2}\right)^3
     - \frac{N_{c} M_t^4}{3 \pi^2 m_{h^0}^2 v^2}  \right\}
      \label{ceff}
\end{eqnarray}
where $M=m_{12}/\sqrt{\sin\beta\cos\beta}$, $M_\Phi^{}$ and $p_i$ represent the mass of the $H^0$, $A^0$ or $H^\pm$ bosons
and the momenta of the external Higgs lines, respectively.
We note that in Eq.~(\ref{ceff}) $m_{h^0}$ is the renormalized
physical mass of the lightest CP-even Higgs boson $h^0$. In our calculation of the cross section of $pp \to h^0h^0$ in the decoupling limit, we ignore one-loop effects due to the $h^0b\bar{b}$ coupling and replace the $h^0h^0h^0$ coupling by its effective coupling given in Eq.~\ref{ceff}.
%----------------------------------------
%-------------------------------------------------------------------
%-------------------------------------------------------------------
\begin{figure}[h!]
\centering
\includegraphics[width=8.1cm]{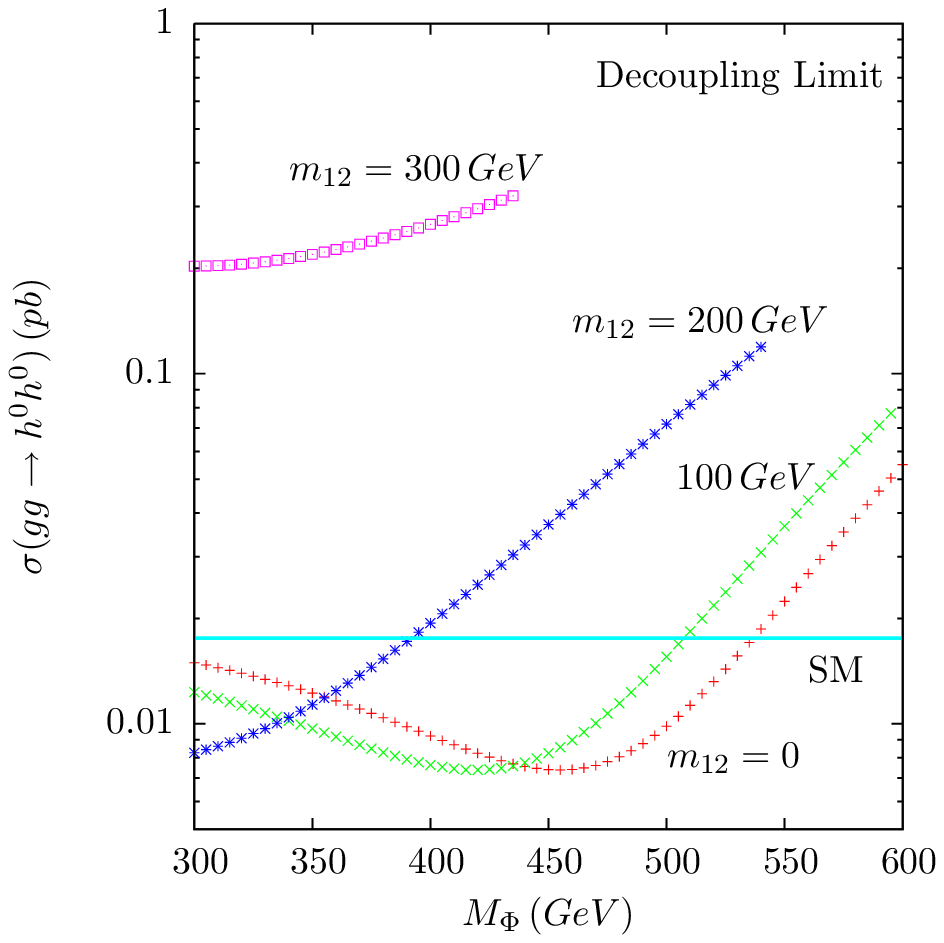}
\includegraphics[width=7.8cm]{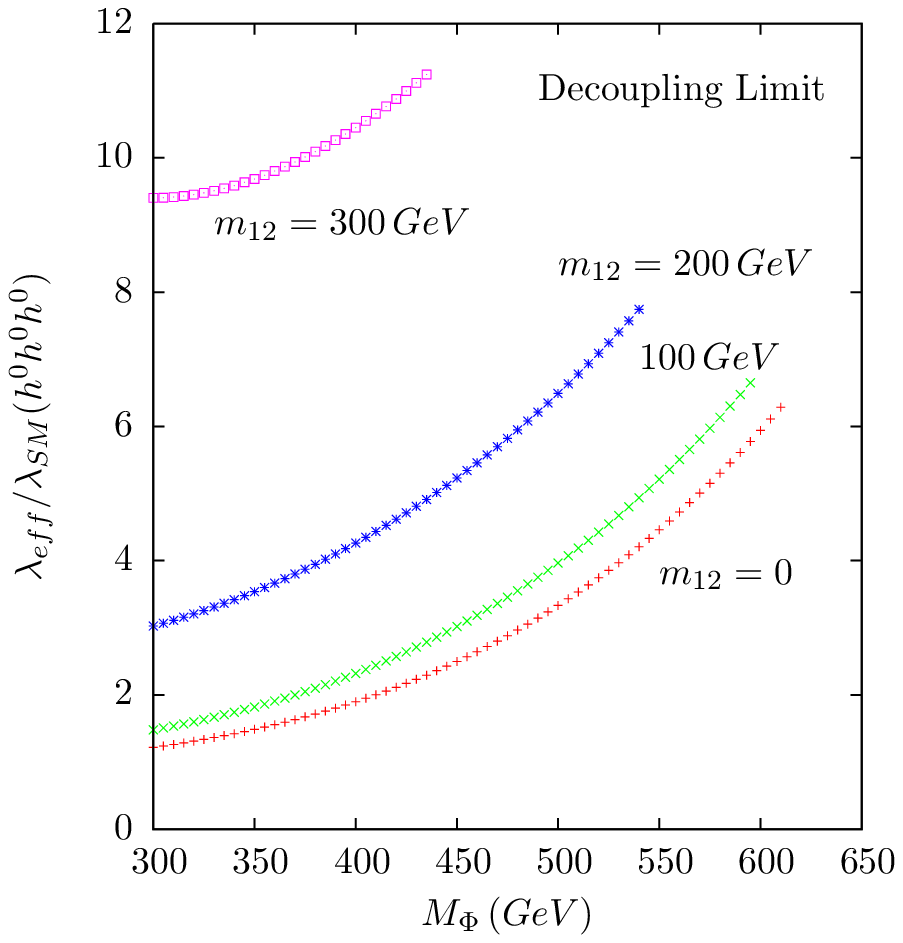}
\caption {$pp \to h^0 h^0$ in the
decoupling limit. On the right we show the strength of the one-loop corrected $h^0h^0h^0$ coupling
relative to SM, for several values of $m_{12}$. On the left plot we show the production cross section for  $m_{h^0}=115$ GeV as a function of $ M_{\Phi}$ for the same values of $m_{12}$ as well as the corresponding SM cross section.}
    \label{fig:decoupling}
\end{figure}
%----------------------------------------
%----------------------------------------
%----------------------------------------
In Fig.~\ref{fig:decoupling} we present the double Higgs production process for the LHC in the decoupling limit of the 2HDM. In the left panel we show $\sigma_{gg\to h^0h^0}$ for $m_{h^0}=115$ as a function of $M_\Phi$ for several values of $m_{12}$. In the right panel we show how the effective coupling behaves with respect to the corresponding SM coupling when we vary $M_\Phi$ and $m_{12}$. Let us first note that in the decoupling limit, the cross section depends only on these three parameters and very mildly on $\tan \beta$ - the sensitivity of cross section to $\tan\beta$ is mild since $\tan\beta$ only enters the $h^0h^0h^0$ coupling through $M^2=m_{12}^2/(\sin\beta\cos\beta)$.  Moreover, $\tan\beta$ is constrained by perturbative unitarity to be rather moderate. The abrupt cuts in the plots are due perturbative unitarity and vacuum stability constraints. We are considering the case $m_{12}^2 > 0$ where the non-decoupling effects are larger.
It is interesting to note that in the decoupling limit the cross sections can
be smaller than the SM one. This is because the box diagrams do not depend on
$m_{12}$ neither on $m_{\Phi}$. For small $m_{\Phi}$, there is a destructive
interference between boxes and vertices but as $m_{\Phi}$ grows the total
contribution becomes dominated by the vertex diagrams which explain why it is
enhanced for large $m_{\Phi}$. If $m_{12}$ is very large (300 GeV or more) the
box contributions become negligible and non-decoupling effects can be seen
already for small $M_\Phi$. For our choice of $m_{h^0}=115$ GeV, the SM cross
section for LHC is small, $0.02$ $pb$.  As is clear in
Fig.~\ref{fig:decoupling}, the 2HDM contributions can enhance the cross
section by about one order of magnitude and reach $0.4$ $pb$ for $m_{12} = 300
\, GeV$. Contrary to what happens in the corresponding process $\gamma \gamma \to h^0
h^0$~\cite{photon}, we will not be able to test non-decoupling effects arising from the $h^0 H^+ H^-$ vertex at one loop level because charged Higgs bosons do not couple to gluons. Only non-decoupling effects from higher order corrections can be probed. In the right panel of Fig.~\ref{fig:decoupling}, one can see that the coupling $h^0h^0h^0$ is enhanced by a factor of more than eight in some cases. This enhancement in the $h^0h^0h^0$ coupling is due to the non decoupling effect shown in Eq.~(\ref{ceff}). Finally we remark that the $b\bar{b}$ contribution is much smaller than the gluon fusion one and has a much higher uncertainty coming from the PDFs. We have also checked that $c\bar{c}\to h^0h^0$ contribution is much smaller that the $b\bar{b}\to h^0h^0$ one.

\subsection{The general case for $pp \to h^0h^0$}

The allowed parameter space of the 2HDM is restricted after imposing experimental and theoretical constraints. Before presenting our results we will scan the parameter space imposing all theoretical constraints discussed before to search for allowed regions where the values of the cross section are above the SM ones. The most severe constraints are the 14 tree-level unitary conditions imposed on the scalar potential parameters.

\begin{figure}[h!]
  \begin{center}
\begin{tabular}{cc}
\hspace{-0.6cm}
\resizebox{83mm}{!}{\includegraphics{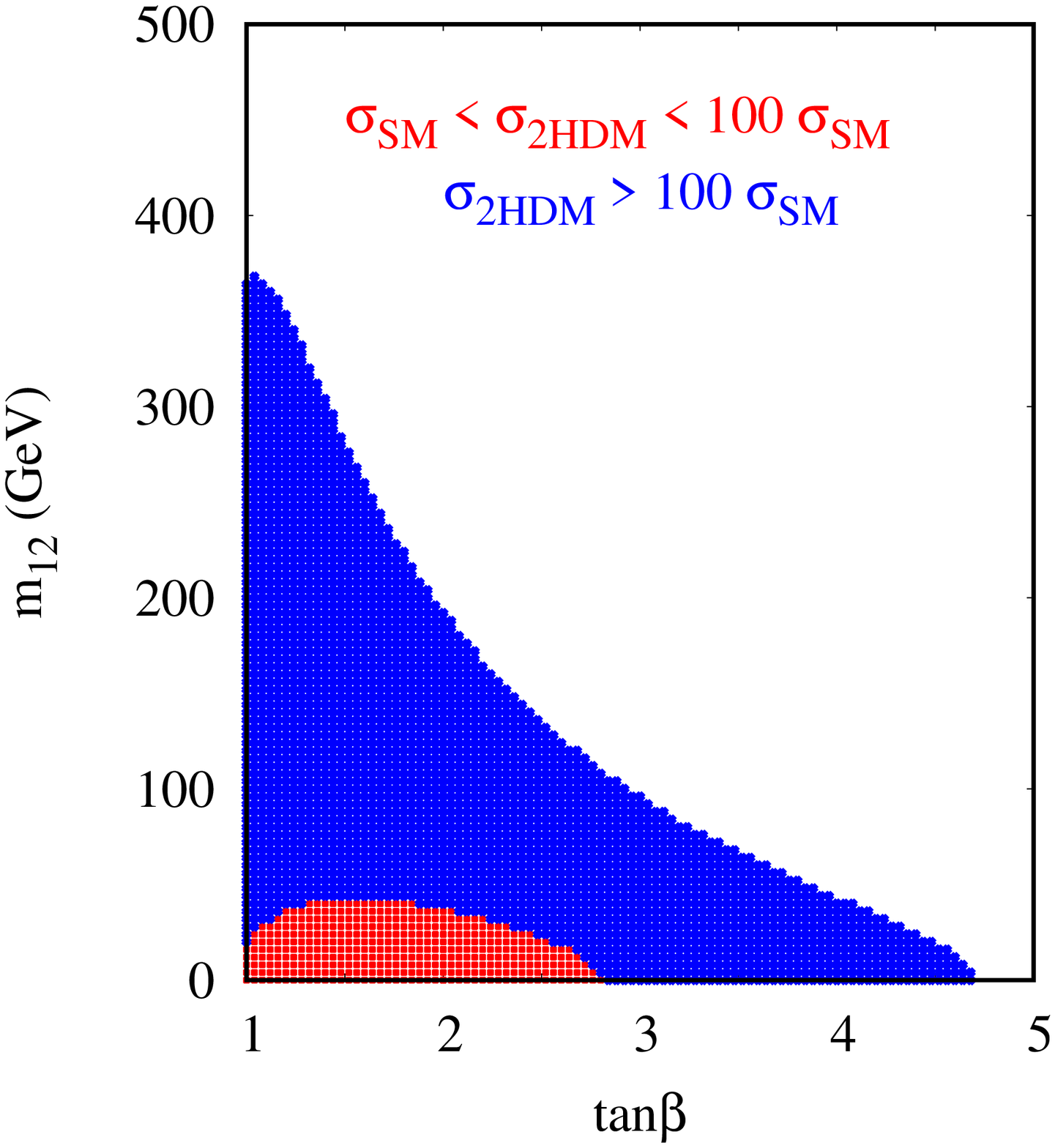}} & \hspace{-0.6cm}
\resizebox{81.5mm}{!}{\includegraphics{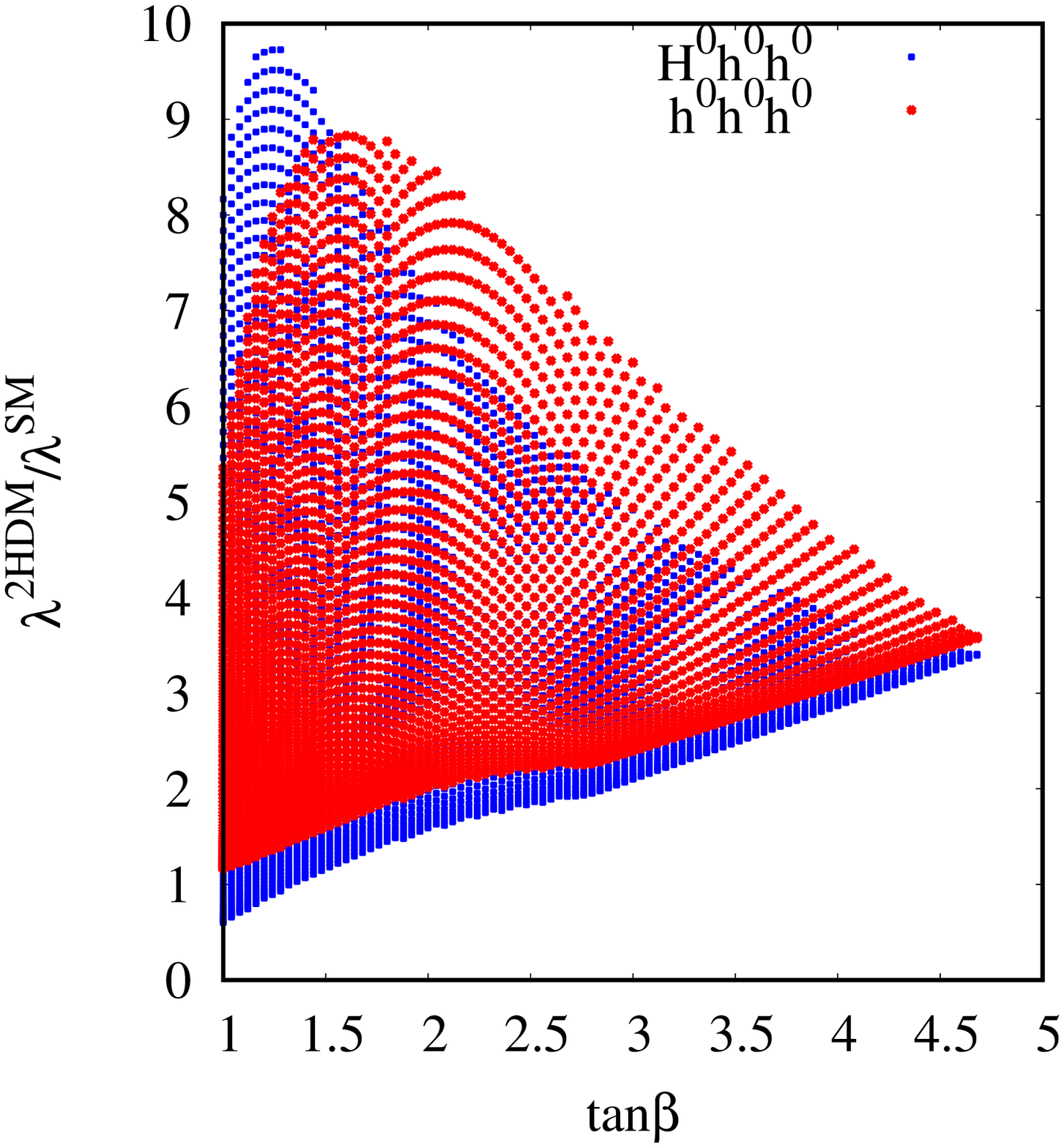}}
\end{tabular}
    \caption{The allowed regions of the 2HDM where the $\sigma(gg \to h^0
      h^0)$ is larger than the corresponding SM cross section (left) and
the 2HDM triple coupling $h^0h^0h^0$ and $h^0h^0h^0$ normalized to SM
couplings (right).  We show $m_{12}$ as a function of $\tan \beta$ for $m_{h^0}= 115$ GeV, $m_{A^0}= 350$ GeV, $m_{H^\pm} = 300 $ GeV, $m_{H^0} = 250 \, GeV$ and $\sin \alpha = - 0.9$.}
    \label{fig:scan2}
  \end{center}
\end{figure}

\begin{figure}[h!]
\begin{tabular}{cc}
\hspace{-0.6cm}
\resizebox{83mm}{!}{\includegraphics{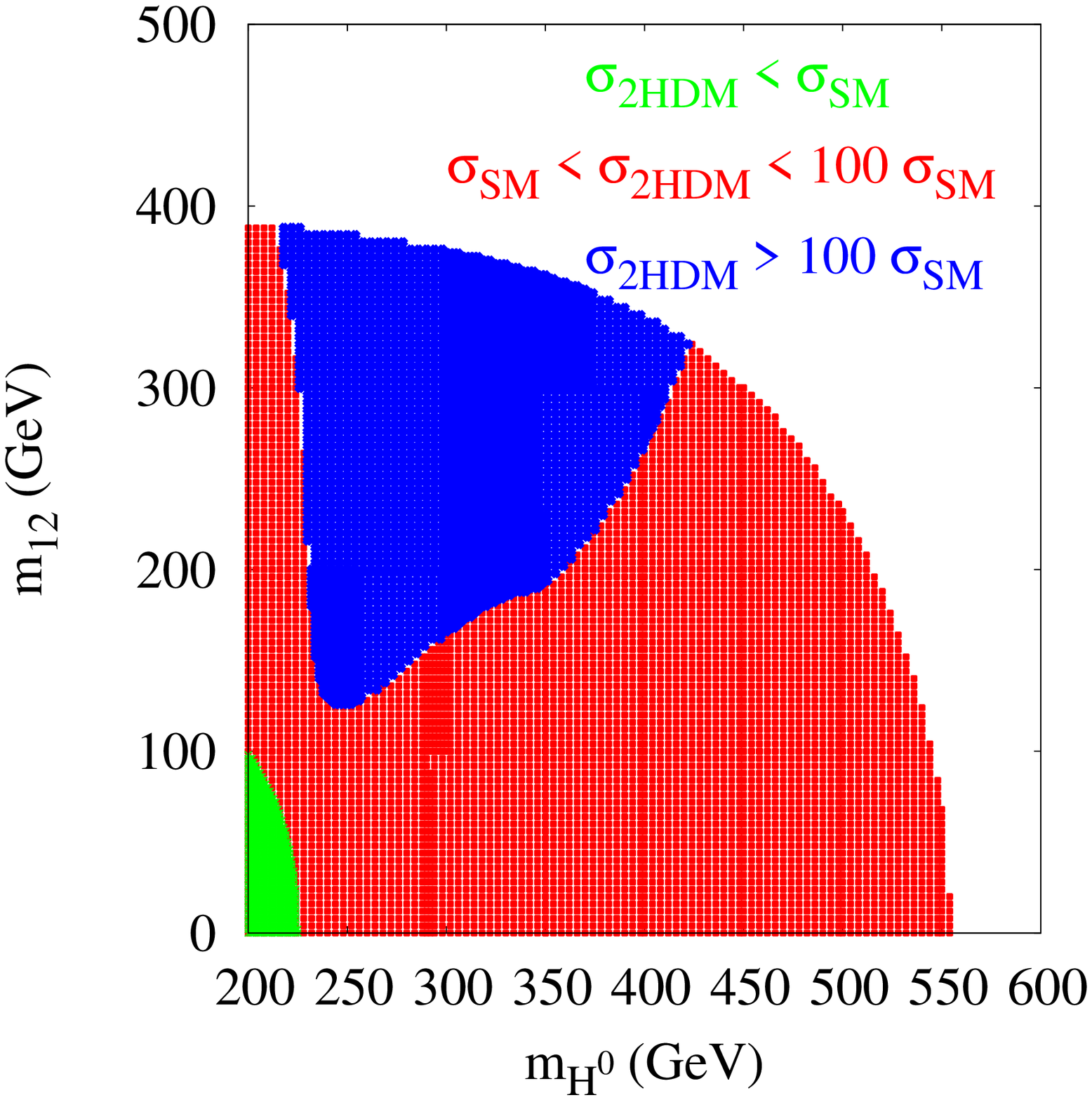}} & \hspace{-0.6cm}
\resizebox{81.5mm}{!}{\includegraphics{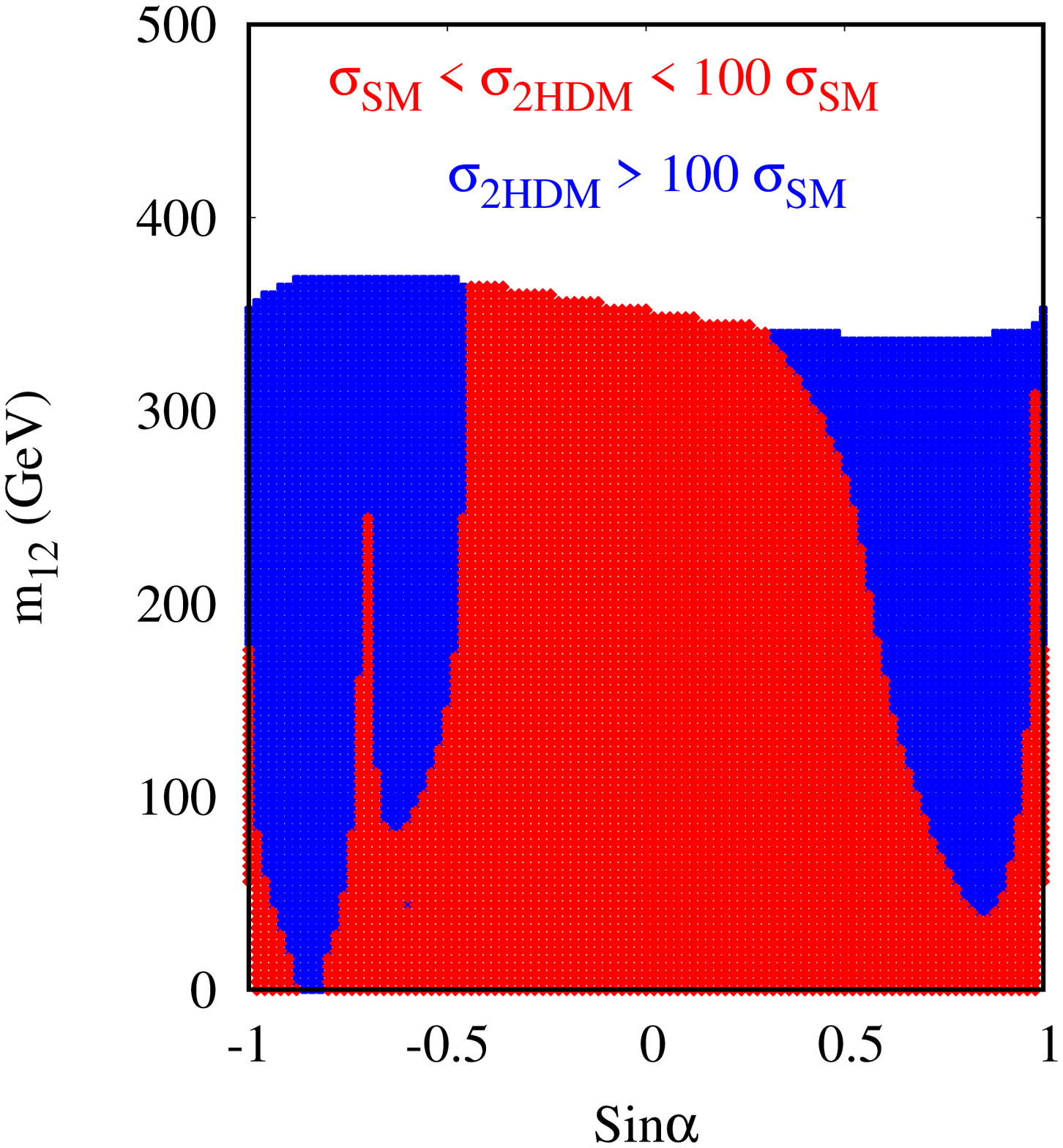}}
\end{tabular}
\caption{
The allowed regions of the 2HDM where the $\sigma(gg \to h^0 h^0)$ is larger than the corresponding SM cross section. We have chosen $m_{h^0}= 115$ GeV, $m_{A^0}= 350$ GeV,
$m_{H^\pm} = 300 $ GeV and $\tan \beta =1$. In the left panel we show $m_{12}$ as a function of $m_{H^0}$ with $\sin \alpha = - 0.9$. In the right panel we plot $m_{12}$ as a function of $\sin \alpha$ for $m_{H^0} = 250 \, GeV$.
}
\label{fig:scan1}
\end{figure}

As explained before, the 2HDM under study has 7 independent parameters. In
this particular study, the mass on the pseudo-scalar, $m_{A^0}$ and the mass
of the charged Higgs, $m_{H^\pm}$, do not play any role in the value obtained
for the cross section. The reason is that the pure scalar couplings
$h^0h^0h^0$,  $H^0H^0H^0$ and $H^0h^0h^0$, Eqs. (\ref{lll}), (\ref{hhh})
and (\ref{hll}), do not involve those parameters. The same is obviously true for
the Yukawa couplings. The only possible contribution from those parameters
occurs when one of the CP-even Higgs is allowed to decay to a pair of
pseudo-scalars or/and to a pair of charged Higgs bosons. In this case, they
will influence the total cross section via a change in the total width of the
CP-even scalar in question. However, in order to check perturbative unitarity
and vacuum stability constraints, $m_{A^0}$ and $m_{H\pm}$ must be specified. The
most relevant parameters we have to consider are therefore $m_{h^0}$,
$m_{H^0}$, $\sin \alpha$, $\tan \beta$ and above all $m_{12}$. We have scanned
the 2HDM parameter space to look for the regions where the cross sections are
larger than the SM one.
The results are shown in Figs.~\ref{fig:scan2} and \ref{fig:scan1}. The detailed parton level studies presented in ~\cite{Baur:2002qd, Baur:2003gp, Baur:2003gpa} concluded that although hard there is a chance of measuring the SM triple Higgs couplings at the LHC. For some final states the measurement could only take place with the tenfold increase luminosity of the Super Large Hadron Collider. Although a more detailed analysis is needed we will take these studies as a benchmark and look for the parameter space where cross sections are at least larger than the SM double Higgs production.
%
%
%-------------------------------------------------------------------------
%-------------------------------------------------------------------------
%
%---------------------------------------------------------------
%-----------------------------------------------------------------

In Fig.~\ref{fig:scan2} we present a scan of $m_{12}$ as a function of $\tan \beta$ for $m_{h^0}= 115$ GeV, $m_{A^0}= 350$ GeV, $m_{H^\pm} = 300$ GeV, $m_{H^0} = 250 \, GeV$, $\tan \beta =1$ and $\sin \alpha = - 0.9$. With this choice of the charged Higgs boson mass, the $b\to s \gamma$ constraint is
fulfilled for all 2HDM models. The scan shows the allowed region of the 2HDM parameter space where the cross section is larger than the SM cross section. Here we see that for reasonable values of the other parameters $\tan \beta$ has to be below 5. We can also witness a correlation between $m_{12}$ and $\tan \beta$: large values of one of the variables force low values of the other variable. In the right panel of Fig.~\ref{fig:scan2}, we illustrate the size of the triple Higgs couplings $h^0h^0h^0$ and $H^0h^0h^0$ normalized to SM
one as a function of $\tan\beta$ and $m_{12}$. For small $\tan\beta$ and large $m_{12}$ the couplings can be larger than $8\times \lambda_{h^0h^0h^0}^{SM}$.

In the left panel of Fig.~\ref{fig:scan1} we again show the allowed region of the 2HDM parameter space where the cross section is larger than the SM cross section. It is clear that large values of $m_{12}$ are preferred because the self-couplings have a strong dependence on this parameter. As expected, when $m_{H^0}\approx 2 m_{h^0}$ the cross section reaches a maximum but all values above $2 m_{h^0}$ can give rise to large cross sections depending on the remaining parameters. From the right  panel of Fig.~\ref{fig:scan1}, where $m_{12}$ as a function of $\sin \alpha$ is shown, we conclude that large values of $|\sin \alpha|$ maximize the cross section. It should be noted however that these plots would change if the values of the masses were modified.

Let us move to the discussion of double Higgs production $\sigma (gg \to h^0
h^0)$ at the LHC in the framework of the general 2HDM. The cases of $H^0 H^0$
and $H^0 h^0$ production will be discussed at the end of this section.
We already know that the cross section does not depend on $m_{A^0}$ nor on $m_{H\pm}$ except via the Higgs $H^0$ or $h^0$ width. We also know that, due to the theoretical constraints, $\tan\beta$ is constrained to be rather moderate. From various scans over the parameters space, we conclude that in order to have large enhancement with respect to SM cross section one needs to take small values of $\tan\beta$ together with large positive $m_{12}^2$. We start by showing how the cross section $pp \to h^0h^0$ behaves with the two CP-even Higgs masses.
%
%------------------------------------------------------------------
\begin{figure}[h!]
\begin{tabular}{cc}
\hspace{-0.6cm}
\resizebox{84mm}{!}{\includegraphics{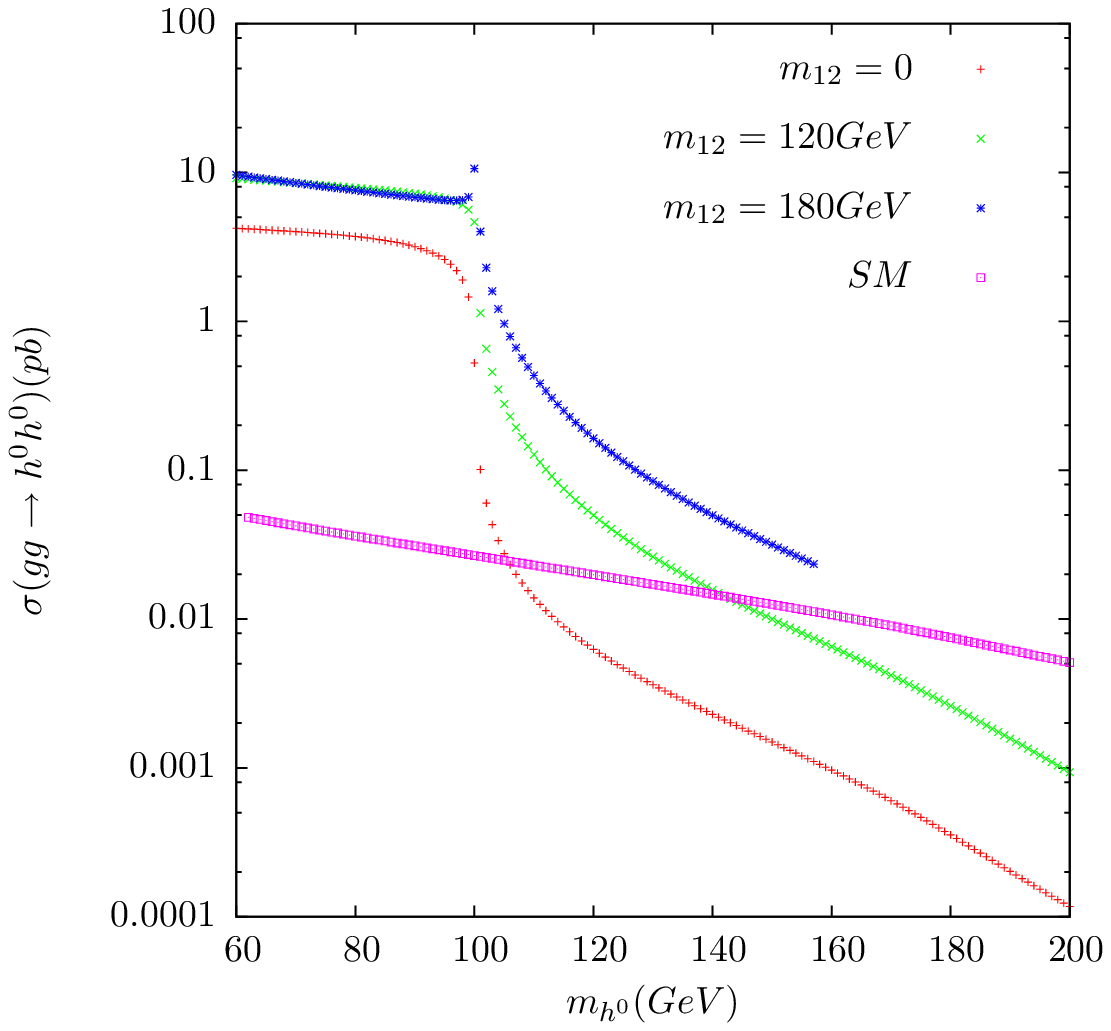}} & \hspace{-0.6cm}
\resizebox{81mm}{!}{\includegraphics{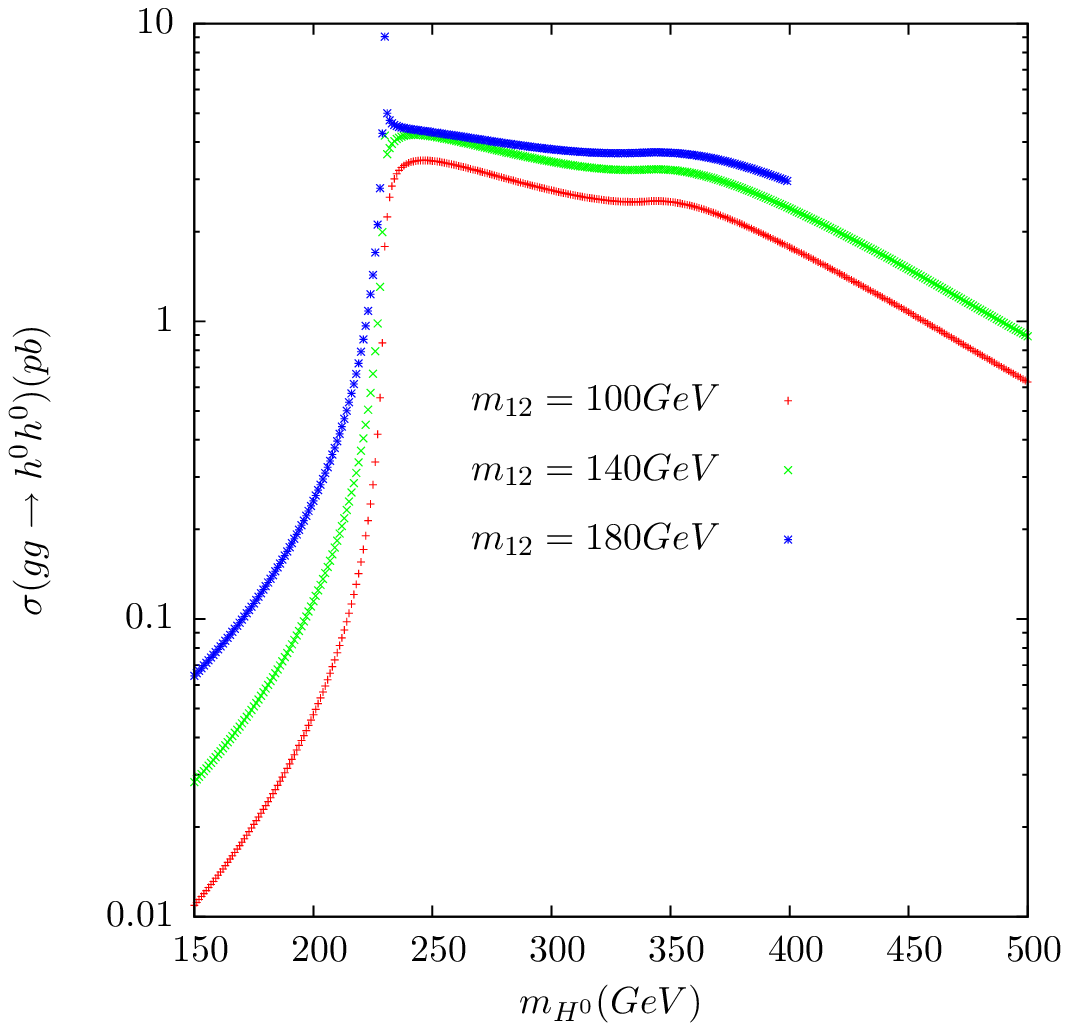}}
\end{tabular}
\caption{The cross section for $gg\rightarrow h^0h^0$ with $m_{A^0} = 320 \, GeV,$ $m_{H^\pm} = 380 \, GeV,$ $\sin (\beta -\alpha) = 0.1$ and $\tan\beta =2$. In the left panel we present the cross section as a function of $m_{h^0}$ for $m_{H^0} = 200 \, GeV$ while in the right panel we can see the variation with $m_{H^0}$ for $m_{h^0} = 120 \, GeV$.}
\label{fig:mhmH}
\end{figure}
%----------------------------------------------------------------
%
This is illustrated in Fig.~\ref{fig:mhmH}, where the cross section $\sigma(gg \to h^0h^0)$ at the LHC via gluon fusion is shown for three different values of $m_{12}$. We have chosen $\sin (\beta - \alpha) = 0.1$ which allow us to probe the very low mass region in $m_{h^0}$. However, we have checked that similar values for the cross section can be obtained for a very wide range of $\sin (\beta - \alpha)$. As expected, the cross section grows with $m_{12}$. In the left panel of Fig.~\ref{fig:mhmH} we show $\sigma(gg \to h^0h^0)$ as a function of $m_{h^0}$ for $m_{H^0} = 200 \, GeV$. Before the threshold, $m_{H^0}=2 m_{h^0}$, the cross section is of the order of a few $pb$ and does not depend much on the value of $m_{12}$. This is because once the channel $H^0 \to h^0 h^0$ is open it becomes dominant with a branching ratio close to $100 \, \%$. On the other hand, there is another triangle diagram contribution, where a virtual $h^0$ is exchanged, with a much smaller coupling constant. Therefore, the dependence on $m_{12}$ is mild. In this region, there is a clear chance of measuring the triple coupling $H^0h^0h^0$ as the cross section can easily be two orders of magnitude above the SM (note that we show the SM line for all values of the light Higgs mass to serve as a reference although masses below 115 GeV are already excluded). Below the threshold, in the region where $H^0 \to h^0 h^0$ is closed, all contributions are now on the same footing and depend much more on the set of parameters chosen. For large enough $m_{12}$ the cross section is still above the SM one but as one can see in Fig.~\ref{fig:mhmH} it can also be well below the SM value for $m_{12}=0$ in the large light Higgs mass region. In the right panel of Fig.~\ref{fig:mhmH} we show a plot for the same set of parameters but now as a function of $m_{H^0}$. This plot show us exactly what we have just discussed but from a different point of view - that of the heavy CP-even Higgs mass. Again we see that there is a limit for what the cross section can grow by increasing $m_{12}$ and that limit is attained when $Br (H^0 \to h^0 h^0) \approx 100 \%$. Finally let us comment that these cross sections do not vary much when we change the Yukawa couplings. There is only a small difference due to the Higgs widths which is however two small to be measured at the LHC.

%------------------------------------------------------
\begin{figure}[h!]
\begin{tabular}{cc}
\hspace{-0.6cm}
\resizebox{84mm}{!}{\includegraphics{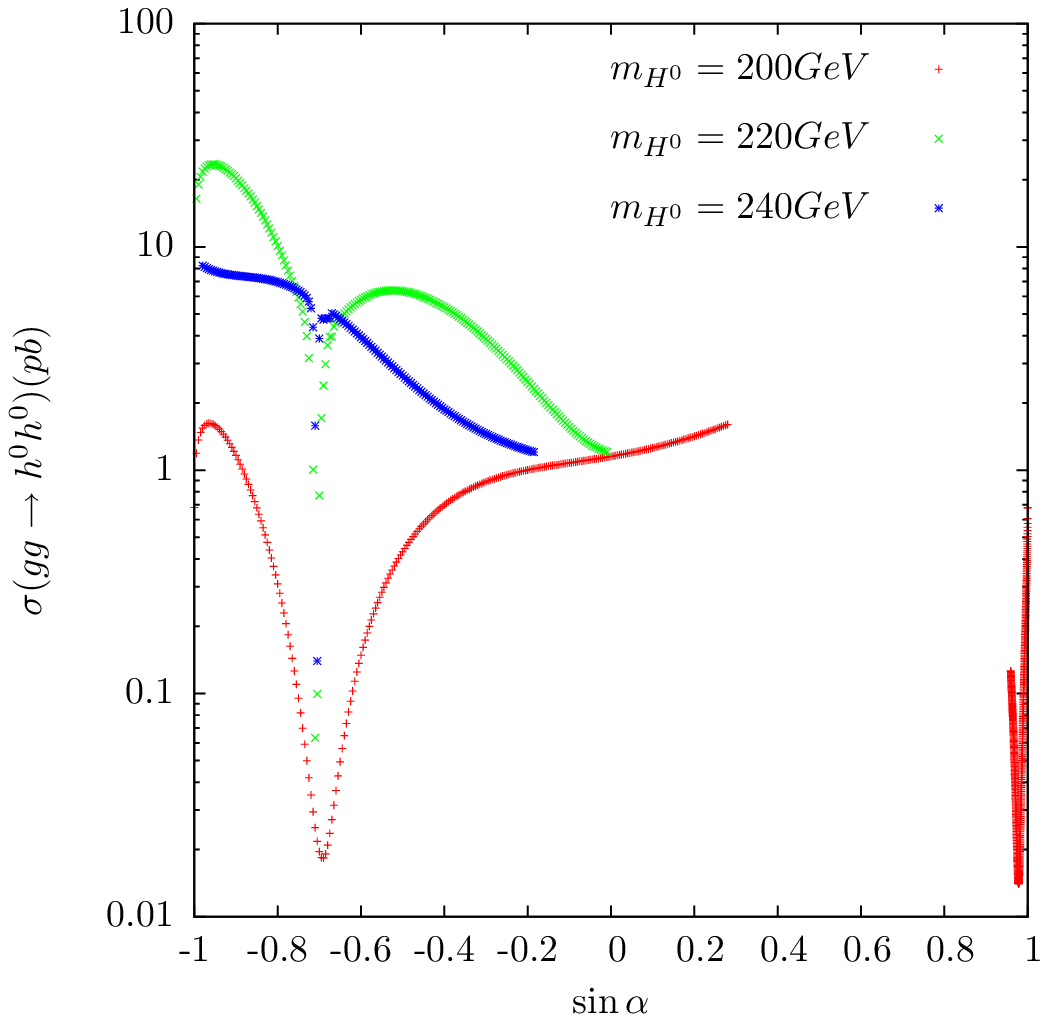}} & \hspace{-0.6cm}
\resizebox{78mm}{!}{\includegraphics{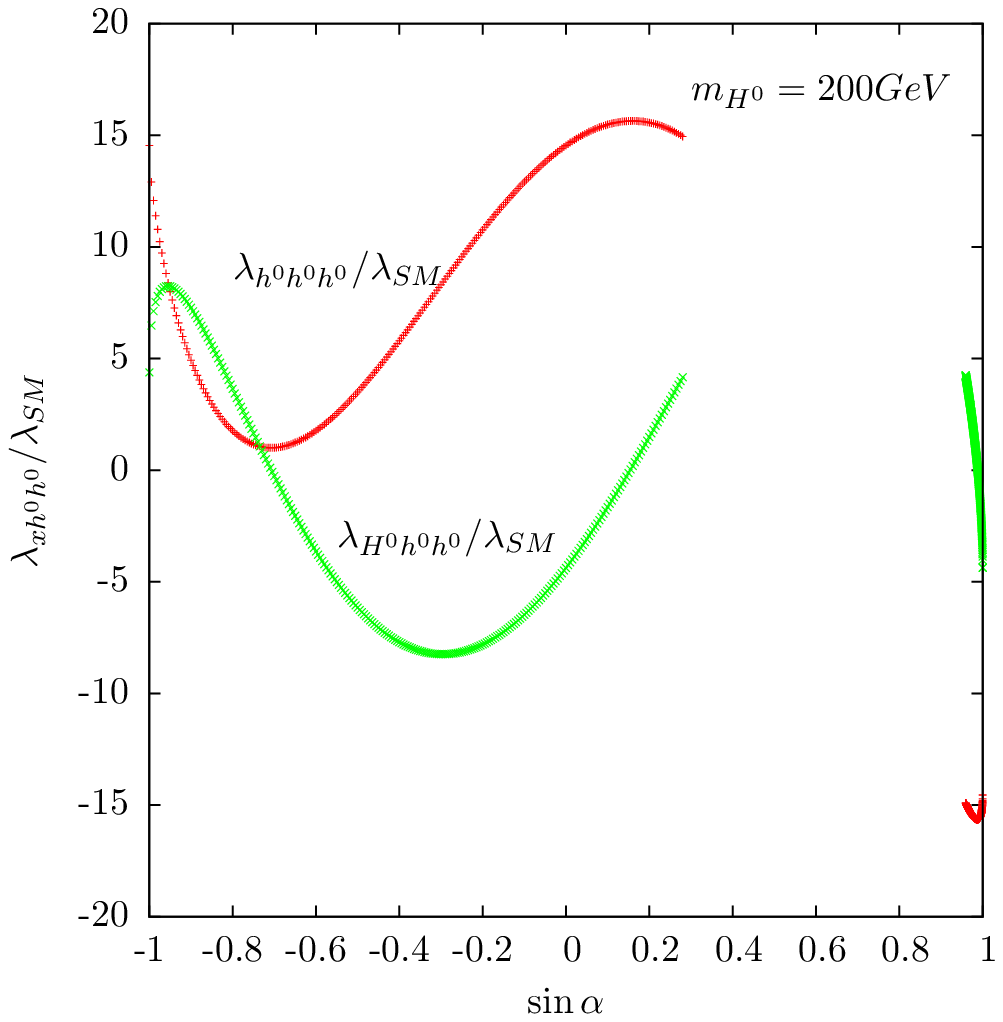}}
\end{tabular}
\caption{The cross section for $gg\rightarrow h^0h^0$ (left) and the contributing scalar couplings (right) as a function of $\sin\alpha$ for different values of $m_{H^0}$. The parameters are $m_{h^0} = 110 \, GeV,$ $m_{A^0} = 320 \, GeV,$ $m_{H^\pm} = 380 \, GeV,$ $m_{12} = 340 GeV,$ and $\tan\beta =1$.}
\label{fig:sinalpha}
\end{figure}
%--------------------------------------------------------

In Fig.~\ref{fig:sinalpha} (left panel) we show the cross section as a
function of $\sin \alpha$ for different values of $m_{H^0}$, one below the
$H^0 \to h^0 h^0$ threshold, one above but almost on the threshold and the
other one above the threshold. In the right panel we present the value of the
couplings $H^0 h^0 h^0$ and $h^0 h^0 h^0$ normalized to SM one
which prove to be the main contributions by far to the value of the cross
section. This was explicitly checked by repeating the calculation with only
the triangle diagrams where either a virtual $h^0$ or a virtual $H^0$ was
exchanged. Combining the information from the two figures we conclude that the
triangle diagram with the virtual $H^0$ exchanged dominates for $-1 < \sin
\alpha < - 0.6$. Therefore in this range, and for this set of parameters the
coupling $H^0 h^0 h^0$ could be measured because the cross section is much
larger than the corresponding SM cross section. After crossing $\sin \alpha =
- 0.6$ both couplings grow and now the relative importance also depends on the
value of $m_{H^0}$: if $H^0 \to h^0 h^0$ is open then the diagram with the
virtual $H^0$ still dominates and there is a good chance of measuring it;
below the threshold, and for this particular set of parameters, it is the diagram with the virtual $h^0$
exchange that dominates. We conclude by saying that depending on the parameters there is a good chance of measuring the coupling $H^0 h^0 h^0$ or $h^0 h^0 h^0$ and that this occurs in a large portion of the parameter space.

\subsection{The $H^0 H^0$ and $H^0 h^0$ production modes}

%-------------------------------------------------------------------------
\begin{figure}[h!]
\begin{tabular}{cc}
\hspace{-0.6cm}
\resizebox{81mm}{!}{\includegraphics{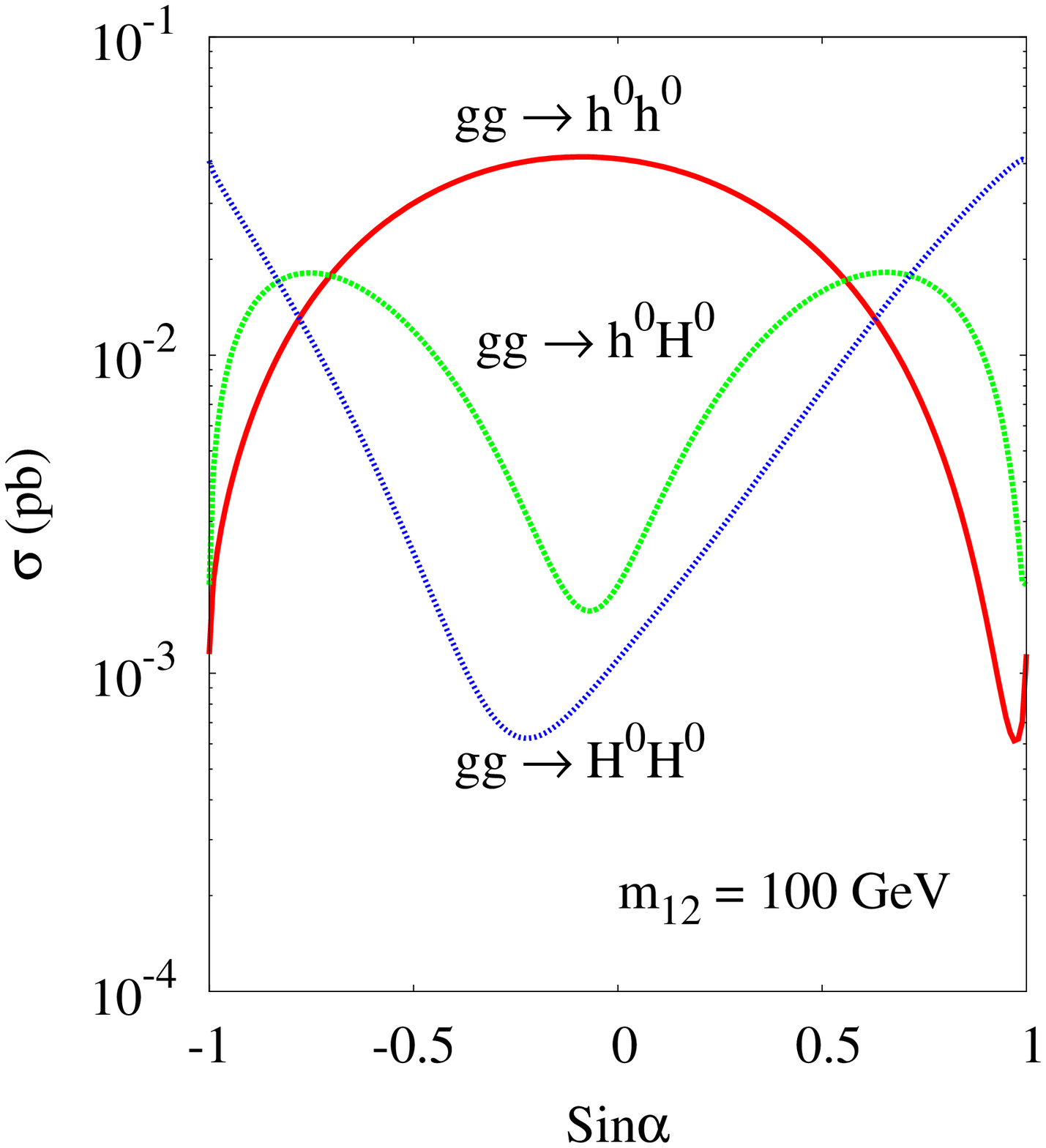}} & \hspace{-0.6cm}
\resizebox{82mm}{!}{\includegraphics{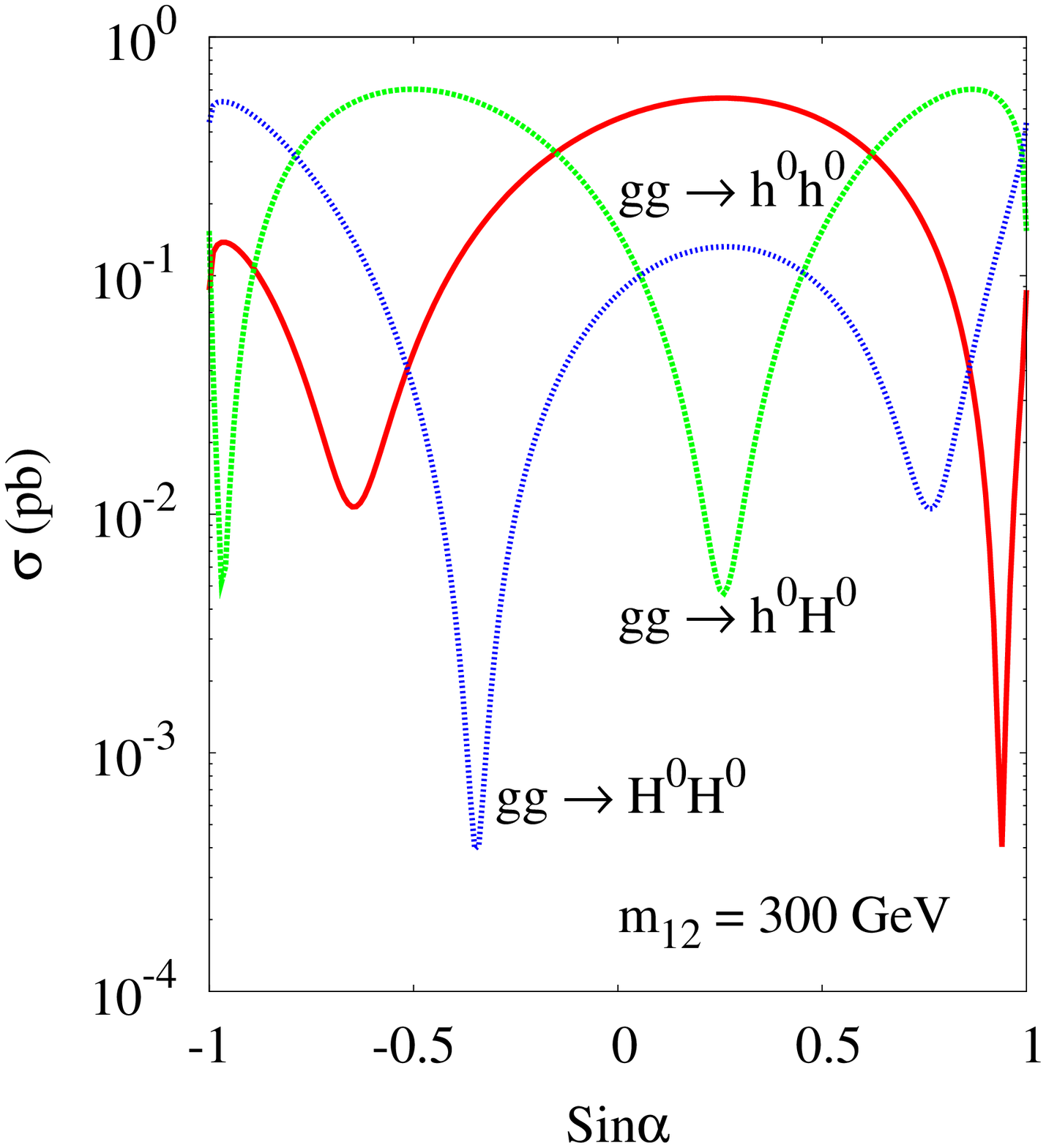}}
\end{tabular}
\caption{$\sigma (gg\rightarrow h^0h^0, \, H^0h^0$ and $ H^0H^0) $ as a function of $\sin\alpha.$ The parameters chosen are $m_{h^0} = 120 \, GeV,$ $m_{H^0} = 121 \, GeV$, $m_{A^0} = 200 \, GeV$, $m_{H^\pm} = 300 \, GeV$ and $\tan\beta =1$. In the left panel $m_{12}=100 \, GeV$ and in the right panel $m_{12}=300 \, GeV$}
\label{fig:sinal}
\end{figure}
%-------------------------------------------------------------------------

Processes $gg \to H^0 H^0, H^0 h^0$ differ from $gg \to h^0 h^0$ mainly due to the fact that $H^0$ is heavier than $h^0$. This means that there is no possibility of resonant behavior - the intermediate state is always off-shell - and that there is a reduction of the phase space available. We may ask ourselves what would the difference be if the two CP-even masses were of similar magnitude. In this case, it is the relation between $\sin \alpha$, $m_{12}$ and $\tan \beta$ that determines the value of the cross section, that is, it is the way these three variables are combined that determines how large the cross section is. Because we know that $\tan \beta$ has to be small we have fixed $\tan \beta=1$ and varied the remaining two parameters for $m_{h^0} = 120 \, GeV$ and $m_{H^0} =121 \, GeV$. In Fig.~\ref{fig:sinal} we present all three cross sections $ \sigma (gg\rightarrow h^0h^0, \, H^0h^0$ and $ H^0H^0) $ as a function of $\sin\alpha$ for $m_{h^0} = 120 \, GeV,$ $m_{H^0} = 121 \, GeV$, $m_A = 200 \, GeV$, $m_{H^\pm} = 300 \, GeV$ and $\tan\beta =1$. In the left panel we take $m_{12}= 100 \, GeV$ while we choose $m_{12}= 300 \, GeV$ for the right panel. We start by noting that we have checked that even if the value of $m_{H^0}$ is shifted up several tenths of $GeV$ the general trends do not change much. It is clear that in any case the cross section can reach several hundreds of fbarn. However, a large $m_{12}$ is needed because for $m_{12}=100 \, GeV$ the cross section barely reaches the SM value. We can see clearly two effects due to the couplings: the huge dependence on the value of $\sin \alpha$ due to the Yukawa couplings and the dependence on the relation between $m_{12}$ and $\sin \alpha$ caused by the scalar couplings. We conclude that if the CP-even scalar masses are similar and not too large, the production of these particles is heavily conditioned by the values of $m_{12}$ and $\sin \alpha$. Finally we should point out that in this particular limit, the two CP-even Higgs bosons would be undistinguishable experimentally and therefore the number of events would be obtained by summing the three contributions.

\section{Higgs signatures}
\label{sec:signatures}

In most of the 2HDM parameter space, the dominant process in the production of a pair of neutral Higgs is the triangle diagram where $gg \to H^0 \to h^0 h^0$. As the mass of the heavy Higgs grows, the triangle diagram $gg \to h^0 \to h^0 h^0$ starts to play an important role. Therefore searches based on these topologies in the SM and in the MSSM should yield similar results. Searches based on $4b$ final state were performed in \cite{Baur:2003gpa} for the SM and in \cite{Dai:1995cb,Dai:1996rn} for the MSSM. The final state $b \bar{b} \tau^+ \tau^-$ was also explored in \cite{Baur:2003gpa} for the SM. The search based on rare decays, $h^0 h^0 \to b \bar{b} \gamma \gamma$ both for the SM and for the MSSM was performed in~\cite{Baur:2003gp}. In this section we will present the profile of the light Higgs boson in all four Yukawa type models. We will then comment on the results, showing that new rare decays can be used in the 2HDM case.
\begin{figure}[h!]
\centering
\includegraphics[height=3.1in]{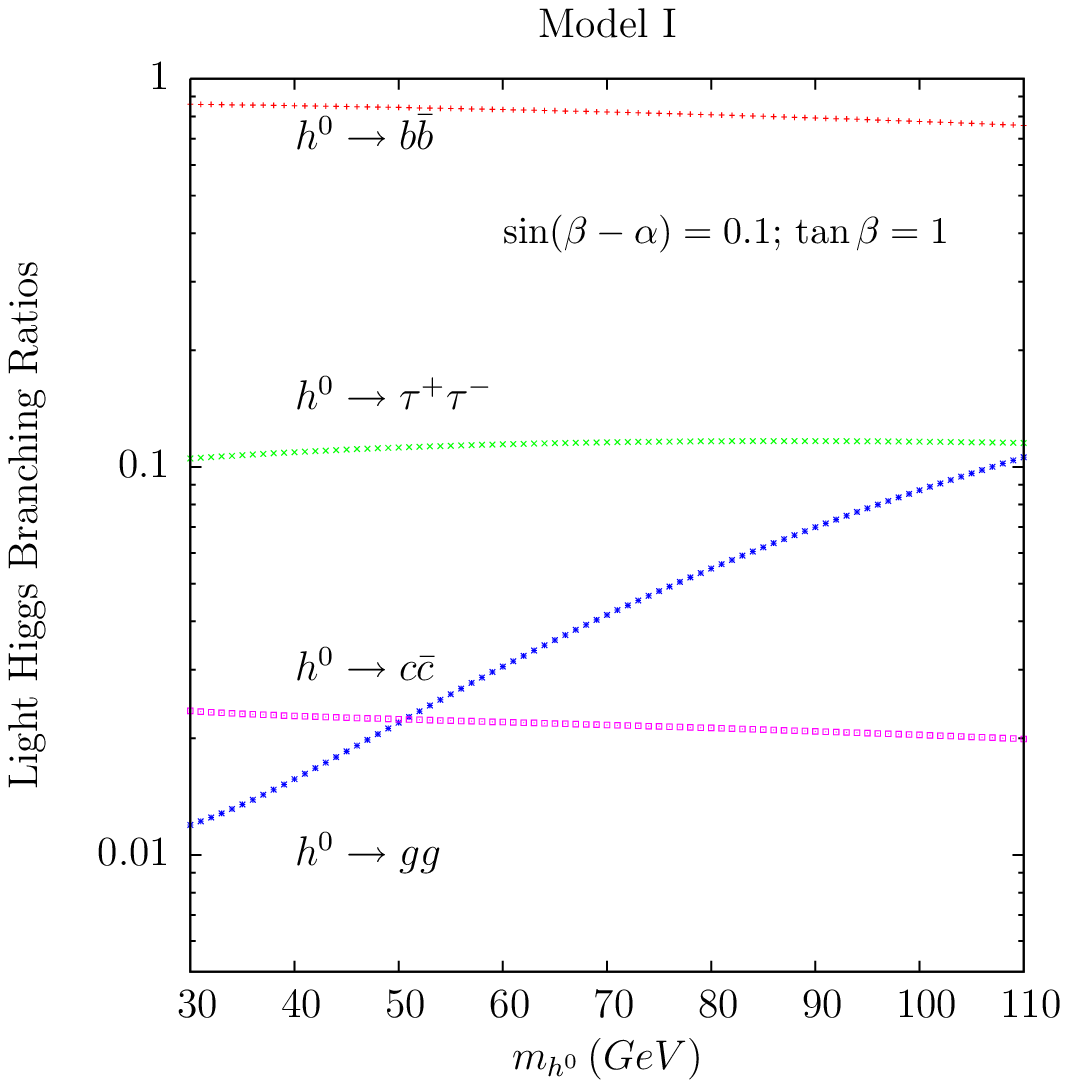}
%\hskip.5cm
\includegraphics[height=3.1in]{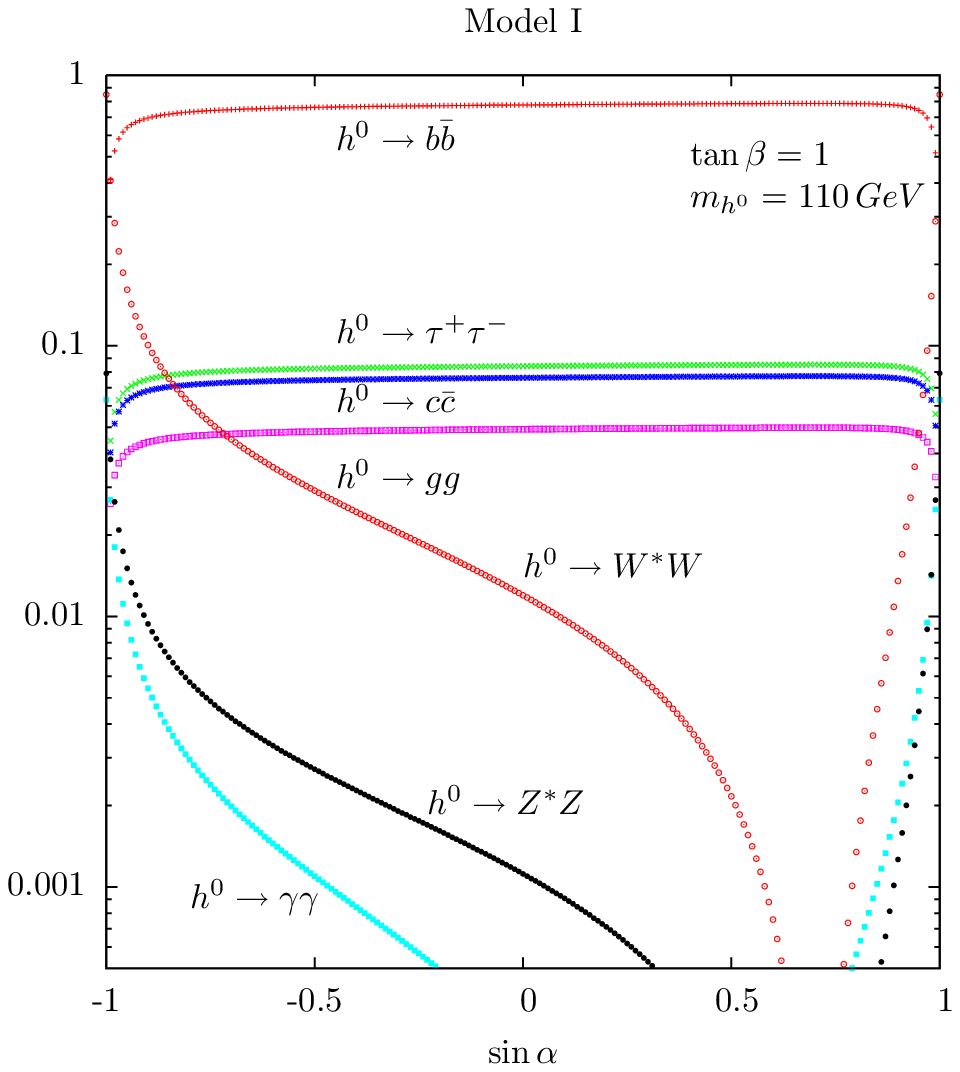}
\caption {Light Higgs branching ratios in model I as a function of its mass  (left panel) and as a function of $\sin \alpha$ (right panel). Although irrelevant in Model I we have chosen $\sin(\beta-\alpha)=0.1$, $\tan \beta = 1$, $m_{H^0}= m_{A^0} = m_{H^\pm}=300$ GeV and $m_{12}=0$ GeV.}
\label{plotlightBR1}
\end{figure}

We will discuss a scenario where the light Higgs boson's mass is well below $ \approx \, 200 \, GeV$ and all decay channels that have at least one other Higgs in the final state are closed. Scenarios where the light Higgs boson decays to a pair of charged Higgs, a pair of CP-odd Higgs or a combination of Higgs and gauge bosons are obviously allowed in a general 2HDM. In the decoupling limit, the lightest CP-even Higgs has the SM signature. This means that close to the limit $(\beta - \alpha) \approx \pi/2$ no major differences should be expected in the light Higgs boson profile with respect to the SM Higgs. The same is true for model type I. In Fig.~\ref{plotlightBR1} we show the profile of the light Higgs boson in Model I as a function of its mass in the left panel and as a function of $\sin \alpha$ in the right panel. As in Model I only one Higgs couples to all fermions, the only dependence is in the mass of the light Higgs. The branching ratios do not depend on any other parameters of the model. It should be noted however that when $\sin \alpha = 1$ (the fermiophobic limit) the decay to gauge bosons become dominant. In this case the light Higgs will decay mainly to two photons or to two W bosons depending on the Higgs mass considered and this trend can be seen in the right panel of Fig.~\ref{plotlightBR1}. The values presented in the right panel, $\sin (\beta - \alpha) = 0.1$ and $\tan \beta = 1$ are irrelevant for Model I for the channels presented. However, they are relevant for the other models and the left panel of Fig.~\ref{plotlightBR1} also shows the qualitative behavior of Models II, III and IV for the same set of parameters.  Therefore, for the set of values presented this plot is the same in all four models. When $\tan \beta = 1$ and at the same time $\beta \approx \alpha$ the angular dependence almost cancels resulting in a SM-like scenario. As $\tan \beta$ grows this is no longer true: in Model II the coupling to up quarks is suppressed resulting again in a SM-like scenario with the $b \bar{b}$ and the $\tau^+ \tau^-$ mode dominating; in Model III all couplings are now suppressed except the ones to the down quarks which means that the Higgs decays almost exclusively to b-pairs; finally in Model IV, and again for high $\tan \beta$ the Higgs decays almost exclusively to $\tau^+ \tau^-$. We should add that, moving away from $\beta \approx \alpha$, as the mass of the light Higgs grows, the decays $h^0 \to W W^*$ and $h^0 \to  Z Z^*$ become increasingly important although their branching ratios are always below the SM values.
\begin{figure}[h!]
\centering
\includegraphics[height=3.1in]{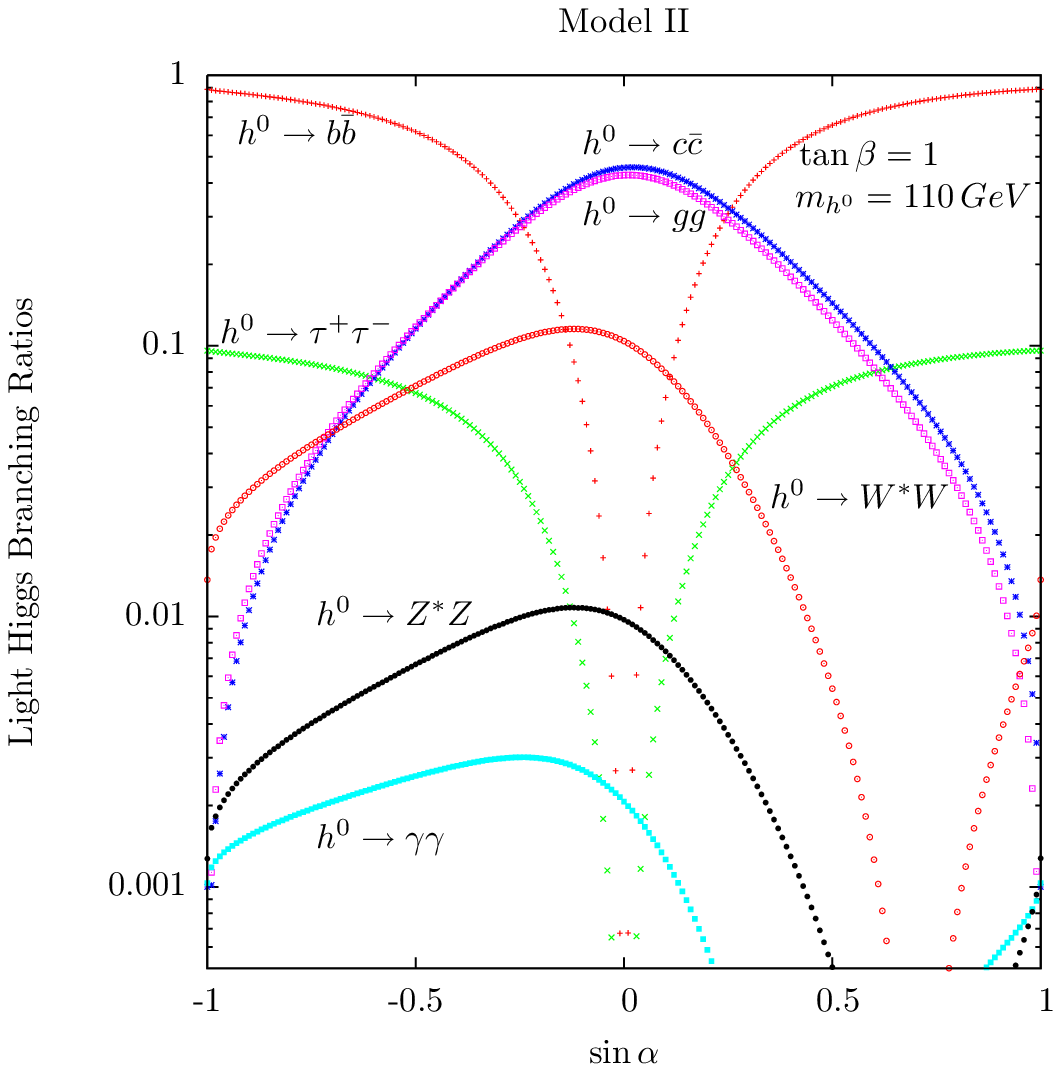}
\includegraphics[height=3.1in]{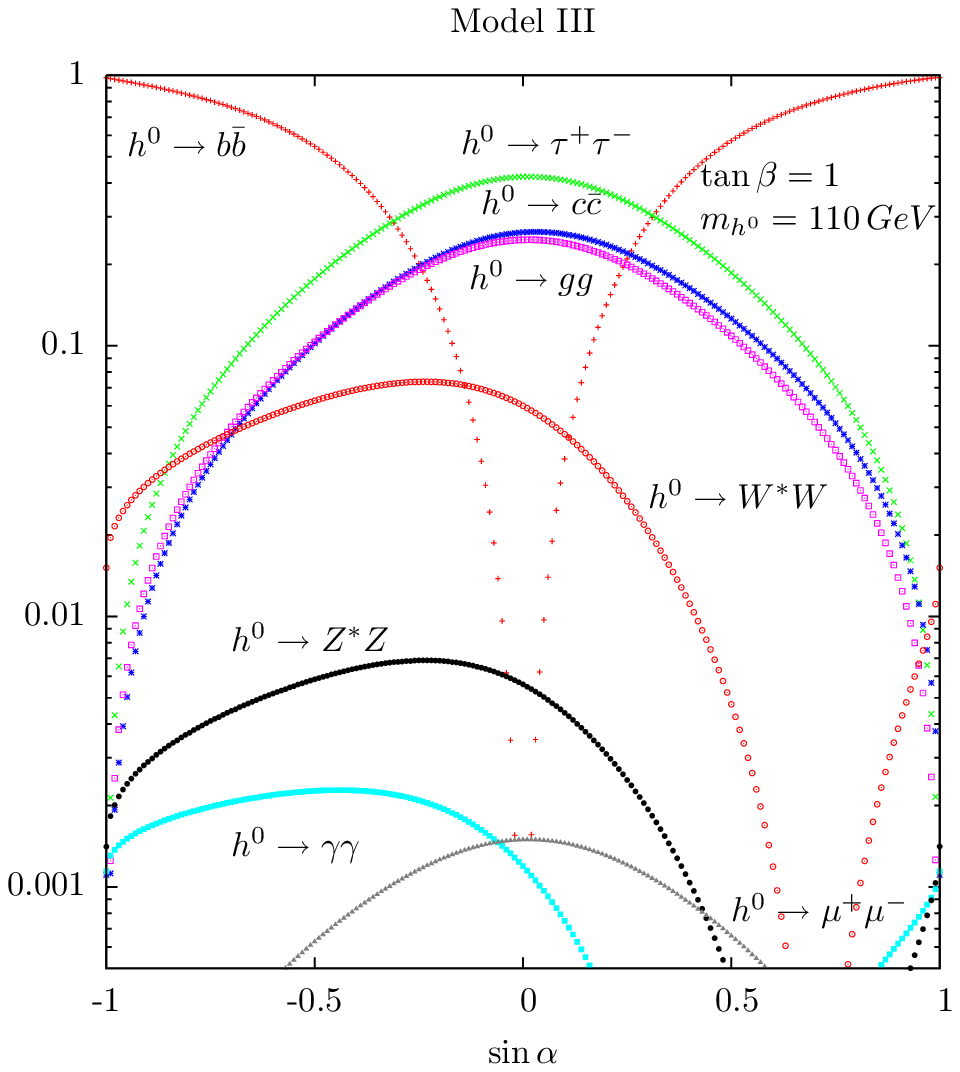}
\caption {Light Higgs branching ratios in models II (left panel) and III (right panel) as a function of $\sin \alpha$ for $\tan \beta = 1$. Again, although irrelevant, we have chosen the other parameters as $m_{H^0}=m_{A^0} = m_{H^\pm}=300$ GeV and $m_{12}=0$ GeV.}
\label{plotlightBR2}
\end{figure}
\begin{figure}[h!]
\centering
\includegraphics[height=3.1in]{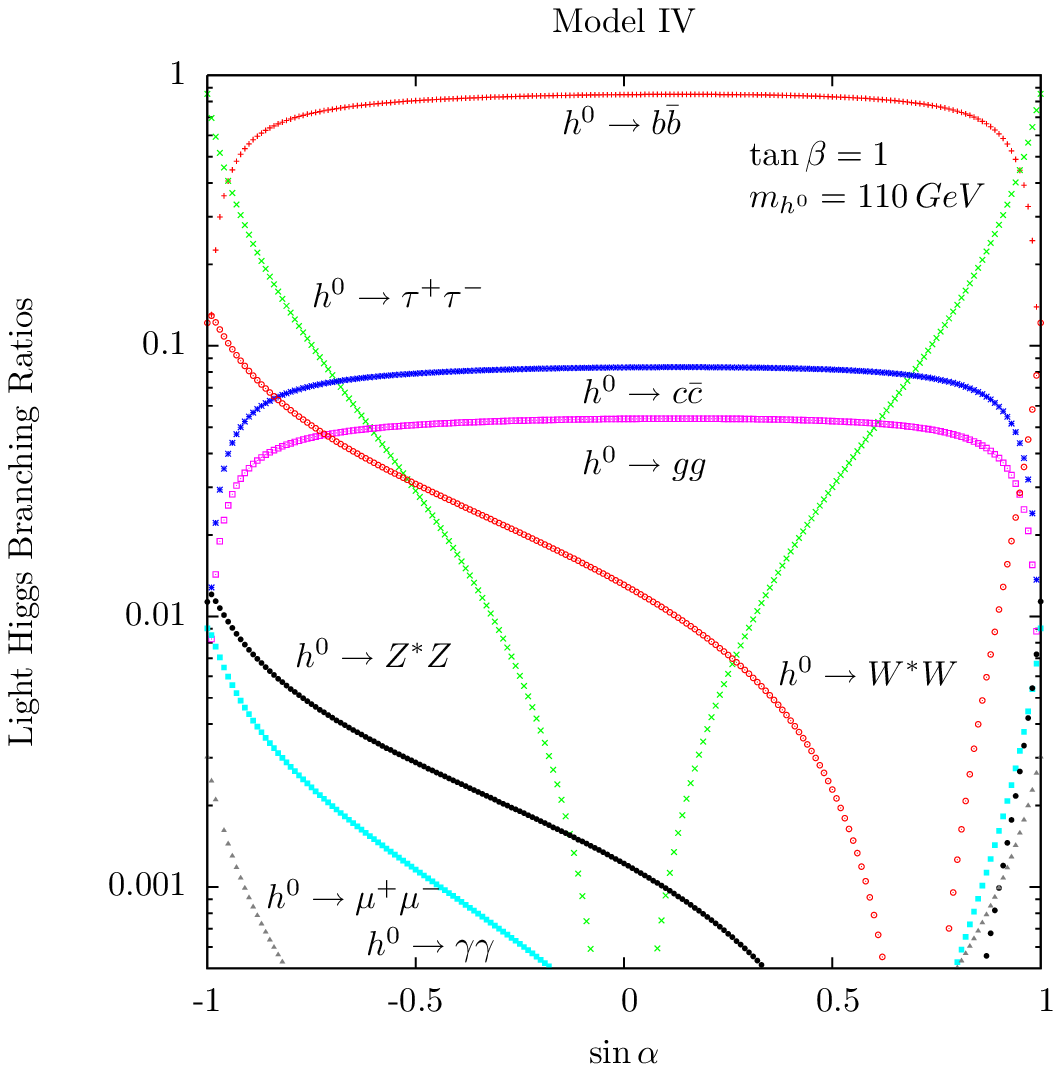}
\includegraphics[height=3.1in]{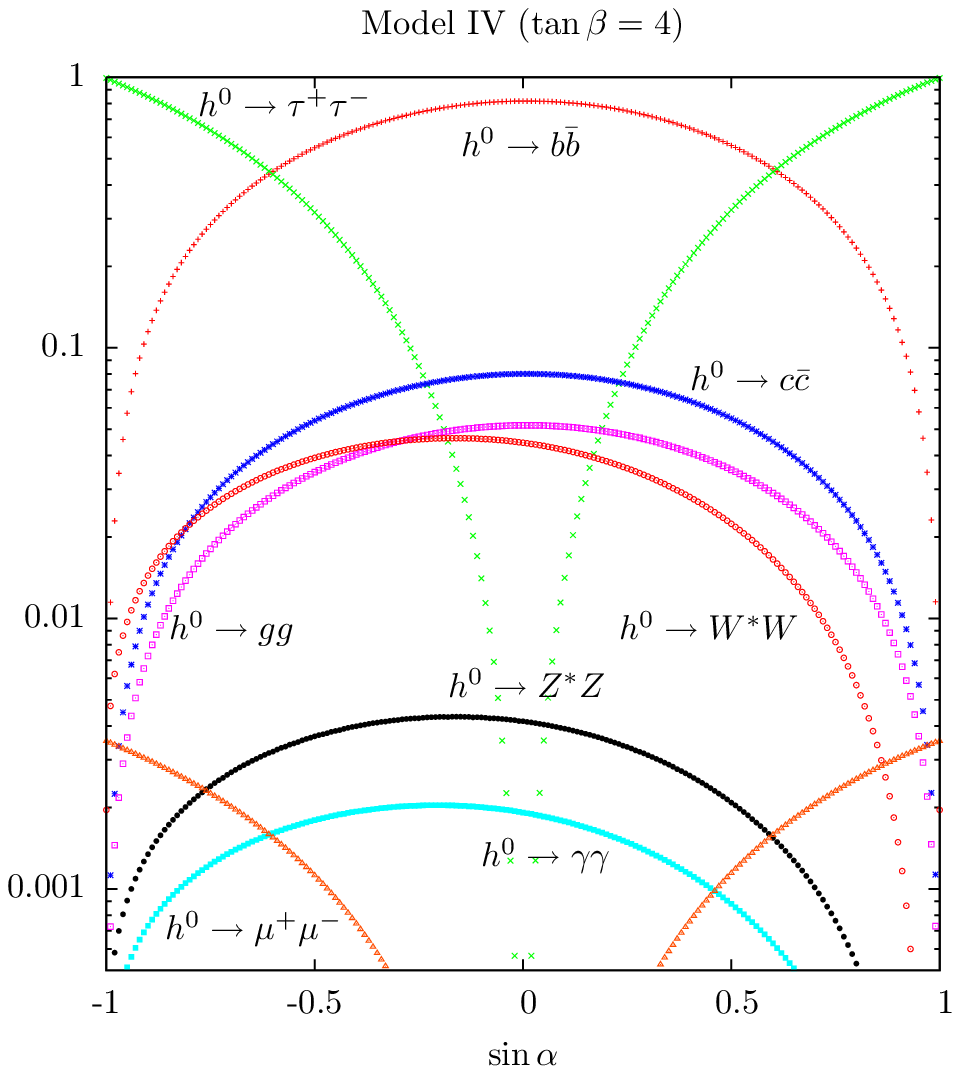}
\caption {Light Higgs branching ratios as a function of $\sin \alpha)$ in model IV for $\tan \beta = 1$ (left panel) and $\tan \beta = 4$ (right panel). The other parameters are $m_{H^0}= m_{A^0} = m_{H^\pm}=300$ GeV and $m_{12}=0$ GeV.}
\label{plotlightBR3}
\end{figure}

Most interesting is the dependence of the light Higgs branching ratio with $\sin \alpha$. In Fig.~\ref{plotlightBR2} we show the light Higgs branching ratios as a function of $\sin \alpha$ for Models II (left) and III (right) and for $\tan \beta = 1$.
In Model II, when $\sin \alpha$ is close to zero, the dominant decays become $c \bar{c}$ and $gg$. Therefore, in this region a decay to non b-jets can easily reach $100 \, \%$ making the detection of Higgs pairs extremely hard due to the overwhelming QCD background. In Model III a more pleasant situation emerges. In the same $\sin \alpha$ region, the decay to $\tau^+ \tau^-$ dominates reaching $50 \, \%$ for $\sin \alpha= 0$. At the same time the decay channel $h^0 \to \mu^+ \mu^-$ is well above the $0.1 \%$ line turning the final state $\tau^+ \tau^- \mu^+ \mu^-$ into a very good candidate. We should point out that these final states occur in significant regions of the parameter space as can be seen in the figures.
Finally, note that we don't show the respective plots for large $\tan \beta$ as we would recover the behavior described previously, that is, in Model II (left panel) the light Higgs would decay exclusively to $b \bar{b}$ and $\tau^+ \tau^-$ and in Model III (right panel) $Br (h^0 \to b \bar{b}) \approx \, 100 \, \%$.

In Fig.~\ref{plotlightBR3} we show the light Higgs branching ratios as a function of $\sin \alpha$ for Model IV and for $\tan \beta = 1$ (left panel) and $\tan \beta = 4$ (right panel). For $\tan \beta = 1 $ the Higgs decays mainly to $b$ pairs. Away from $\sin \alpha = 0$ the decay to $\tau$ pairs can reach interesting values and becomes the dominant decay mode when $|\sin \alpha| \approx 1$. For $\tan \beta = 4$ both modes almost share the total branching ratio between them. Once more we should point out that when $\sin \alpha$ moves away from zero, not only the decay to $\tau$ pairs become the dominant mode, but also the decay to $\mu$ pairs reach almost $0.4 \, \%$ for $|\sin \alpha|=1$. Hence, new decay channels can be explored in the context of 2HDM.

The process $pp \rightarrow H^0 H^0$ gives rise to very
similar signatures. The only disadvantage is that $H^0$ is in principle heavier than $h^0$ and phase space is lost. If the masses are of the same order and below let us say 130 $GeV$, than the main decays are again $b \bar{b}$ and $\tau^+ \tau^-$. For larger values of the mass again the decays $h^0 \to W W^*$ and $h^0 \to  Z Z^*$ become dominant. In this limit they differ only in the behavior with $\sin \alpha$ and $\tan \beta$ in all Yukawa models except model I. Finally, process $pp \rightarrow h^0 H^0,$ has in principle two drawbacks: first
the phase space is again reduced due to the presence of a heavier Higgs in the final state; second we can not reduce the background by asking the two invariant masses from each pair that originates  from a Higgs bosons (for instance two b-jets) to have a similar magnitude.

\section{Discussion and Conclusions}
\label{sec:summary}

In this work we have calculated the cross section for the production processes $pp \to h^0 h^0, \, H^0 H^0, \, H^0 h^0$. We have shown that gluon fusion is always the dominant contribution. Regarding the different Yukawa models, little difference was found in the scanned parameter space - although the cross sections presented are for models II and III they are representative of all Yukawa models.

We started by assessing the possibility of distinguishing the lightest CP-even Higgs boson from the SM Higgs in the decoupling limit of the 2HDM. We concluded that the non-decoupling effects entering the amplitude via higher order corrections to the vertex $h^0 h^0 h^0$ can be seen for very large values of the soft breaking parameter $m_{12}$ or for very large values of the common mass scale $M_\Phi$. Taking for instance $m_{12}=300 \, GeV$ the cross section is more than one order of magnitude above the corresponding SM value.

The general 2HDM has too many parameters to allow for a detailed analysis taking all of them into account. However, for the processes studied the only relevant parameters are $m_{12}$, $\sin \alpha$, $\tan \beta$, $m_{H^0}$ and $m_{h^0}$. The remaining masses have only little influence via their contribution to the CP-even Higgs widths. We have scanned the parameter space to conclude that, as a general trend, the largest cross sections are obtained for large $m_{12}$ and large $|\sin \alpha|$. The parameter $\tan \beta$ is constrained to be $O(1)$ except for very specific scenarios. Note however that we have showed that even for $m_{12} =0$ cross sections can be 100 times above the SM cross section when $m_{H^0}> 2 m_{h^0}$.  The cross sections are larger for smaller values of the CP-even Higgs masses and can be two orders of magnitude above the SM cross section if the channel $H^0 \to h^0 h^0$ is open. We have shown that there are scenarios where the triple couplings could be measured.

We have then moved to the study of the other production modes $H^0 H^0$ and $H^0 h^0$. We have showed that all production modes can be large and that the main variation occurs with the combination of $\sin \alpha$ and $m_{12}$ parameters. In fact, we have shown that when $m_{H^0} \approx m_{h^0}$ the different processes differ only by virtue of values of the parameters $\sin \alpha$ and $m_{12}$ chosen as $\tan \beta$ is bounded to be small.

In the last section we have studied the possible final states for each production mode. We have shown that the main decays for a light CP-even Higgs are to b-quark pairs or to $\tau^+ \tau^-$. The relative importance of each of these decays depend very much on the values of $\sin \alpha$ and $\tan \beta$. In~\cite{Baur:2003gpa} it was shown that the best channel to measure the SM triple Higgs coupling is $2b \, 2 \gamma$ for a Higgs mass below 140 $GeV$. In all Yukawa versions of the 2HDM, the branching ratio of Higgs to two photons does not differ  much from the SM. For most of the parameter space, the decay to $b \bar{b}$ is either equal or enhanced relative to the SM. Therefore, this channel should again be one of the most promising to measure 2HDM triple couplings. The channel $2b \mu^+ \mu^-$ was also studied and shown to be one order of magnitude below $2b \, 2 \gamma$. With larger values for the production cross section and the enhancement of the $h^0 \to \mu^+ \mu^-$ branching ratio in some of the parameter space of Models III and IV, this is a channel worth exploring. Finally,  other channels like $4b$, $2b \tau^+ \tau^-$, and $2b \mu^+ \mu^-$ could be reevaluated for the low mass region when the 2HDM cross section is well above the SM one. For heavier Higgs we would have two distinguish two scenarios. One, where decay channels with at least one other Higgs in the final state are open; in this scenario a new study would have to be performed~\cite{Kanemura:2009mk}. The other where the decays to $WW^{(*)}$ and $ZZ^{(*)}$ become important, which means a larger $h^0$ mass and $\alpha - \beta$ away from zero; in this scenario, the analysis will heavily depend on the values of the branching ratios of the decays $h^0 \to WW^{(*)}$ and $h^0 \to ZZ^{(*)}$.

%-------------------------------------------------------------
\section{acknowledgments}
A. A. acknowledges the warm hospitality of the University of Southampton, where some of this research
took place, and financial support for a visit there from the Science and Technology Facilities Council in
the form of a `Visiting Researcher Grant'. A.A is supported by the National Science of Theoretical Studies-Taipei under contract \# 980528731. R.B. is
supported by National Cheng Kung University Grant No. HUA
97-03-02-063.
C.C.H is supported by the National Science Council of
R.O.C under Grant \#s: NSC-97-2112-M-006-001-MY3. R.B.G. is supported by a Funda\c{c}\~ao para a Ci\^encia e Tecnologia Grant SFRH/BPD/47348/2008. R.S. is supported by the FP7 via a Marie Curie Intra European Fellowship, contract number PIEF-GA-2008-221707.
%--------------------------------------------------------------

\appendix

\section{Fermion-scalar vertices}

\begin{table}[h]
\begin{center}
\begin{tabular}{c c c c c c c c } \hline \hline
        &  \textbf{I} && \textbf{II} && \textbf{III} && \textbf{IV} \\ \hline
                 \hline
$\alpha_{eh}$ & $-\frac{\cos \alpha}{\sin \beta}$ && $\frac{\sin
\alpha}{\cos \beta}$ && $-\frac{\cos \alpha}{\sin \beta}$ &&
$\frac{\sin \alpha}{\cos \beta}$ \\ \hline
$\alpha_{dh}$ & $-\frac{\cos \alpha}{\sin \beta}$ && $\frac{\sin
\alpha}{\cos \beta}$ && $\frac{\sin \alpha}{\cos \beta}$ &&
$-\frac{\cos \alpha}{\sin \beta}$
 \\ \hline
$\alpha_{eH}$ & $-\frac{\sin \alpha}{\sin \beta}$ && $-\frac{\cos
\alpha}{\cos \beta}$ && $-\frac{\sin \alpha}{\sin \beta}$ &&
$-\frac{\cos \alpha}{\cos \beta}$ \\ \hline
$\alpha_{dH}$ & $-\frac{\sin \alpha}{\sin \beta}$ && $-\frac{\cos
\alpha}{\cos \beta}$ && $-\frac{\cos
\alpha}{\cos \beta}$ &&  $-\frac{\sin \alpha}{\sin \beta}$ \\
\hline
$\beta_{e}$ & $-\cot \beta$ && $\tan \beta$ && $-\cot \beta$ &&
$\tan \beta$ \\ \hline
$\beta_{d}$ & $-\cot \beta$ && $\tan \beta$ && $\tan \beta$ && $-
\cot \beta$ \\ \hline \hline
\end{tabular}
\caption{Coupling constants for the fermion-scalar interactions}
\end{center}
\end{table}

\begin{tabular}{lclclll}
$\overline{e}_i e_i h$ & & $\frac{ig}{2 M_W}\alpha_{eh} m_{e_i}$ &
$\qquad \qquad \qquad \overline{u}_i u_i G_0$ & & $-\frac{g}{2 M_W} m_{u_i}
\gamma_5$ \\ \\
$\overline{u}_i u_i h$ & & $-\frac{ig}{2 M_W}\frac{\cos
\alpha}{\sin \beta} m_{u_i}$ &
$\qquad \qquad \qquad\overline{d}_i d_i G_0$ & & $\frac{g}{2 M_W} m_{d_i}
\gamma_5$ \\ \\
$\overline{d}_i d_i h$ & & $\frac{ig}{2 M_W}\alpha_{dh} m_{d_i}$ &
$\qquad \qquad \qquad\overline{e}_i \nu_i H^+$ & & $\frac{ig}{2 \sqrt
{2}M_W}\beta_{e} m_{e_i} (1 + \gamma_5)$ \\ \\
$\overline{e}_i e_i H$ & & $\frac{ig}{2 M_W}\alpha_{eH} m_{e_i}$ &
$\qquad \qquad \qquad\overline{u}_i d_j H^+$ & & $\frac{ig}{2 \sqrt{2}M_W} V_
{ij} \left[ \beta_{d} m_{d_j} (1 + \gamma_5) + \cot \beta m_{u_i} (1 -
\gamma_5) \right]$ \\ \\
$\overline{u}_i u_i H$ & & $-\frac{ig}{2 M_W}\frac{\sin
\alpha}{\sin \beta} m_{u_i}$ &
$\qquad \qquad \qquad\overline{\nu}_i e_i H^-$ & & $\frac{ig}{2 \sqrt
{2}M_W}\beta_{e} m_{e_i} (1 - \gamma_5)$ \\ \\
$\overline{d}_i d_i H$ & & $\frac{ig}{2 M_W}\alpha_{dH} m_{d_i}$ &
$\qquad \qquad \qquad\overline{d}_i u_j H^-$ & & $\frac{ig}{2 \sqrt{2}M_W} V_
{ij}^* \left[ \beta_{d} m_{d_i} (1 - \gamma_5) + \cot \beta m_{u_j} (1 +
\gamma_5) \right]$ \\ \\
$\overline{e}_i e_i A$ & & $-\frac{g}{2 M_W}\beta_{e} m_{e_i}
\gamma_5$ &
$\qquad \qquad \qquad\overline{e}_i \nu_i G^+$ & & $-\frac{ig}{2 \sqrt{2}M_W}
m_{e_i} (1 + \gamma_5)$ \\ \\
$\overline{u}_i u_i A$ & & $-\frac{g}{2 M_W} \cot \beta m_{u_i}
\gamma_5$ &
$\qquad \qquad \qquad\overline{u}_i d_j G^+$ & & $\frac{ig}{2 \sqrt{2}M_W} V_
{ij} \left[ -m_{d_j} (1 + \gamma_5) + m_{u_i} (1 - \gamma_5) \right]$ \\ \\
$\overline{d}_i d_i A$ & & $-\frac{g}{2 M_W}\beta_{d} m_{d_i}
\gamma_5$ &
$\qquad \qquad \qquad\overline{\nu}_i e_i G^-$ & & $-\frac{ig}{2 \sqrt{2}M_W}
m_{e_i} (1 - \gamma_5)$ \\ \\
$\overline{e}_i e_i G_0$ & & $\frac{g}{2 M_W} m_{e_i} \gamma_5$&
$\qquad \qquad \qquad\overline{d}_i u_j G^-$ & & $\frac{ig}{2 \sqrt{2}M_W}V_
{ij}^* \left[ -m_{d_i} (1 - \gamma_5) + m_{u_j} (1 + \gamma_5) \right]$ \\ \\
\end{tabular}

\end{document}